# 3D cine-magnetic resonance imaging using spatial and temporal implicit neural representation learning (STINR-MR)

Hua-Chieh Shao, Tielige Mengke, Jie Deng, You Zhang

*The Medical Artificial Intelligence and Automation (MAIA) Laboratory*
*Department of Radiation Oncology, University of Texas Southwestern Medical Center, Dallas, TX 75390, USA*

## Abstract

**Objective:** 3D cine-magnetic resonance imaging (cine-MRI) can capture images of the human body volume with high spatial and temporal resolutions to study the anatomical dynamics. However, the reconstruction of 3D cine-MRI is challenged by highly undersampled k-space data in each dynamic (cine) frame, due to the slow speed of MR signal acquisition. We proposed a machine learning-based framework, <u>s</u>patial and <u>t</u>emporal <u>i</u>mplicit <u>n</u>eural <u>r</u>epresentation learning (STINR-MR), for accurate 3D cine-MRI reconstruction from highly undersampled data.

**Approach:** STINR-MR used a joint reconstruction and deformable registration approach to achieve a high acceleration factor for cine volumetric imaging. It addressed the ill-posed spatiotemporal reconstruction problem by solving a reference-frame 3D MR image and a corresponding motion model which deforms the reference frame to each cine frame. The reference-frame 3D MR image was reconstructed as a spatial implicit neural representation (INR) network, which learns the mapping from input 3D spatial coordinates to corresponding MR values. The dynamic motion model was constructed via a temporal INR, as well as basis deformation vector fields (DVFs) extracted from prior/onboard 4D-MRIs using principal component analysis (PCA). The learned temporal INR encodes input time points and outputs corresponding weighting factors to combine the basis DVFs into time-resolved motion fields that represent cine-frame-specific dynamics. STINR-MR was evaluated using MR data simulated from the 4D extended cardiac-torso (XCAT) digital phantom, as well as MR data acquired clinically from a healthy human subject. Its reconstruction accuracy was also compared with that of the model-based non-rigid motion estimation method (MR-MOTUS).

**Main results:** STINR-MR can reconstruct 3D cine-MR images with high temporal (<100 ms) and spatial (3 mm) resolutions. Compared with MR-MOTUS, STINR-MR consistently reconstructed images with better image quality and fewer artifacts and achieved superior tumor localization accuracy via the solved dynamic DVFs. For the XCAT study, STINR reconstructed the tumors to a mean±S.D. center-of-mass error of 1.0±0.4 mm, compared to 3.4±1.0 mm of the MR-MOTUS method. The high-frame-rate reconstruction capability of STINR-MR allows different irregular motion patterns to be accurately captured.

**Significance:** STINR-MR provides a lightweight and efficient framework for accurate 3D cine-MRI reconstruction. It is a 'one-shot' method that does not require external data for pre-training, allowing it to avoid generalizability issues typically encountered in deep learning-based methods.

## 1. Introduction

Magnetic resonance imaging (MRI) is a non-invasive modality that can capture morphological and functional characteristics to detect and diagnose health problems (Constantine et al. 2004, Bartsch et al. 2006, Frisoni et al. 2010, Dregely et al. 2018), and to provide image guidance for treatment planning and interventions (Cleary et al. 2010, Pollard et al. 2017, Stemkens et al. 2018), without utilizing ionizing radiation. With advances in hardware designs and innovations in pulse sequences for efficient k-space sampling, time-resolved cine-MRI can now be acquired to visualize time-varying dynamic processes (Nayak et al. 2022), including cardiac motion (Rajiah et al. 2023), blood perfusion (Jahng et al. 2014), speech and vocal production (Lingala et al. 2016), and gas flow in lungs (Wild et al. 2003), etc. However, due to the still-limited speed of k-space data sampling, many of the current cine-MRI applications are limited to 2D, although 3D cine-MRI is highly desired to capture the complex motion/deformation of anatomical volumes (Seppenwoolde, Shirato et al. 2002) to guide diagnosis or treatments, such as the MR-guided radiotherapy (Corradini et al. 2019, Hall et al. 2019, Witt et al. 2020). Considering a pulse sequence with a 4-millisecond (ms) repetition time (TR), for a 100-ms temporal resolution there are only 25 available k-space readout lines (i.e., frequency encoding) to reconstruct a 3D cine-MRI, which is extremely undersampled. Recently, 4D-MRI was developed (Stemkens et al. 2018, Curtis et al. 2022), by repeatedly measuring the dynamic processes and then retrospectively sorting the acquired MR data into 'motion bins' to capture a nominal, averaged motion cycle. The repeated measurements help to secure sufficient data to overcome the undersampling challenge of the dynamic reconstruction problem. However, the repeated measurements and the subsequent motion sorting implicitly assume that the underlying anatomical motion is regular and reproducible, which usually does not reflect the clinical reality (Yasue et al. 2022). Irregular motion patterns can result in degraded image quality (blurriness, ghosting, and other motion artifacts). Also, the averaged motion curves of 4D-MRIs cannot represent such irregular motion that can be important to determine the appropriate radiotherapy margin size or to accumulate the delivered radiation dose. Therefore, reconstructing dynamic 3D cine-MRI is highly desirable in clinics, but remains a challenging problem to solve due to extreme undersampling.

In the past decades, substantial efforts have been put into developing reconstruction algorithms for undersampled k-space measurements. These reconstruction algorithms can be categorized into two main types: model-based iterative algorithms and learning-based techniques (Ravishankar et al. 2020). The first type relies on parallel imaging (Hamilton et al. 2017) and compressed sensing, as well as the corresponding system models (Lustig et al. 2007, Feng et al. 2017). Parallel imaging uses the spatial information from the sensitivity profiles of phased array coils to remove aliasing MRI artifacts or to recover missing k-space data. Compressed sensing regularizes the sparsity of MR images in transformed domains to aid image reconstruction using incoherent measurements. For time-resolved MRI, the spatiotemporal correlation was further exploited to balance the temporal resolution, the spatial resolution, and the image quality (Tsao et al. 2003, Jung et al. 2009, Uecker et al. 2010, Asif et al. 2013, Feng et al. 2016). However, the acceleration factors in these model-based algorithms remain limited (typically $\leq 10$), and compressed sensing-based regularization may lead to overly-smoothed, blurred images under extreme undersampling scenarios (Jaspan et al. 2015). Moreover, these algorithms were mostly driven by non-linear iterative optimization and thus computationally demanding, leading to lengthy reconstruction durations. Accordingly, these methods were largely limited to 2D reconstructions with small numbers of cine frames. To achieve higher acceleration factors to enable 3D cine-MRI reconstruction, deformable image registration was introduced to replace the traditional reconstruction approaches. These deformation-based algorithms reconstructed dynamic MRI frames by estimating the motion fields of underlying subjects with respect to a reference MR image, based on limited-sampled k-space data. The reference MR image was reconstructed either from a separate MR scan or from a subset of the dynamic MR acquisitions. In particular, Huttinga et al. developed a framework, MR-MOTUS, for model-based non-rigid motion and dynamic 3D MRI estimation (Huttinga et al. 2020, Huttinga et al. 2021, Huttinga et al. 2022). Via MR-MOTUS, a 250-frame 3D cine-MRI can be reconstructed with 30 readout lines per frame. However, the accuracy of pure deformation-driven techniques like MR-MOTUS is susceptible to the quality of the reference image. If a separate scan was used to acquire the reference image, the non-deformation intensity variations between the reference image and the dynamic MR acquisition will impact the deformation accuracy (Zhang et al. 2017). If a subset of the dynamic MR acquisitions is used to reconstruct the reference image, the accuracy will instead be impacted by the aliasing artifacts (from undersampling) and/or the motion artifacts (from intra-subset motion) of the reference image.

The second type of techniques are leaning-based, particularly DL-based techniques (Liang et al. 2020). Schlemper et al. developed a cascaded network to unroll the reconstruction problem into joint reconstruction and DL-based de-aliasing (Schlemper et al. 2018). To facilitate the learning of spatiotemporal features for dynamic reconstruction, they introduced data sharing layers and demonstrated an 11-fold acceleration for 2D dynamic cardiac MRI. Biswas et al. introduced a DL framework that incorporated prior information for image denoising, including patient-specific smoothness regularization on a manifold prior and a deep learned prior (Biswas et al. 2019). The algorithm can reconstruct a 200-frame 2D cardiac MRI with 10 readout lines per frame. Huang et al. proposed a motion-guided network comprised of three sub-networks for initial image reconstruction, motion estimation, and motion compensation (Huang et al. 2021), which showed an 8-fold acceleration for 2D cardiac MRI. Although these DL-based methods demonstrated impressive results in cine-MRI reconstruction, the majority of these studies focused on 2D reconstructions as 3D cine-MRI reconstruction is challenged by more extreme undersampling. Similar to the scenario of the first-type algorithms, the deformation-based approaches were also introduced into DL-based frameworks, which can potentially achieve real-time 3D cine-MRI with a high acceleration factor and low inference latency (Terpstra et al. 2021, Shao et al. 2022). However, similar to the deformation-driven algorithms like MR-MOTUS, the DL-based algorithms are impacted by the non-deformation intensity variations between the reference image and the dynamic MR acquisition, or the aliasing/motion artifacts of the reference image. Another major drawback of these DL-based techniques is the model uncertainty and the lack of robustness. The DL-based techniques need to be partially or fully pre-trained, and any data distribution shifts between training and testing can lead to generalizability issues and substantially degrade their accuracy (Zech et al. 2018, Kelly et al. 2019, Full et al. 2021).

In addition to the above DL-based methods, recently a new machine learning technique, implicit neural representation (INR), has found potential applications in medical image reconstruction, registration, and analysis (Khan et al. 2022, Molaei et al. 2023, Rao et al. 2023). INR uses neural networks to implicitly represent physical features of objects (e.g., geometry and material properties such as opacity, x-ray attenuation coefficient, or MR intensity) in a complex 3D scene (Mildenhall et al. 2022, Tewari et al. 2022). A neural network in INR functions as a universal function approximator (Hornik et al. 1989) which takes spatial coordinates of a scene (MR image voxel coordinates, for instance) as inputs and continuously maps them to the desired physical features (MR intensities at the queried voxels) via the learning process. The implicit representation via networks allows the underlying MR image to be captured compactly without specifying the function form in advance (Tewari et al. 2022), and allows natural super-resolution since the MR image intensity can be queried at arbitrary, non-integer coordinates (Chen et al. 2022). In contrast to DL-based methods, which typically require a large dataset for pre-training, INR can be trained in a single shot by directly using limited samples of the studied subject to optimize the network parameters. Therefore, INR is learning efficient and can avoid the generalizability issues typically encountered in DL-based techniques. With these advantages, INR has been applied to solve x-ray-based and MR-based reconstruction problems from sparse-view measurements (Shen et al. 2022, Zha et al. 2022). Furthermore, INR-based reconstruction algorithms for dynamic computed tomography (CT) and cone-beam CT were also developed (Reed et al. 2021, Zhang et al. 2023).

Inspired by our recent work in INR-based cone-beam CT reconstruction (Zhang et al. 2023), in this work we proposed a joint reconstruction and deformable registration-based framework using spatial and temporal INRs for dynamic 3D cine-MRI reconstruction (STINR-MR). STINR-MR uses spatiotemporal INRs to learn, reconstruct, and map 3D cine-MR volumes and the corresponding time-varying motion. It reconstructs a reference-frame image and solves time-varying motion fields with respect to the reference frame to derive corresponding 3D cine-MR images. Compared with pure deformation-driven methods like MR-MOTUS, STINR allows simultaneous reconstruction and motion modeling to solve/optimize the reference-frame image directly and iteratively from the cine k-space data, and thus is not affected by the non-deformation variations between the reference MR image and the cine-MR images. The reconstruction/optimization of the reference MR image using the full cine k-space data also renders it less susceptible to the aliasing/motion artifacts. In contrast to our prior STINR work, we used a powerful learning-based input encoding scheme (multiresolution Hash encoding) for STINR-MR, rendering it a light-weight and efficient framework capable of reconstructing 3D cine-MRIs of >1,000 frames within a short duration (≲ 20 min). STINR-MR was evaluated by MR data simulated from a 4D extended cardiac-torso (XCAT) digital phantom (Segars et al. 2010) featuring various regular/irregular breathing patterns. It was also evaluated by the MR data of a healthy human subject from a publicly available repository (Huttinga et al. 2021). The reconstruction and motion tracking accuracy of STINR-MR was also compared with that of MR-MOTUS.

## 2. Materials and methods

2.1 Problem formulation

Let $\{w_t(k)\}_{t=0}^{N_t-1}$ be a series of consecutive 3D MR acquisitions in k-space, where $w_t(k)$ denotes the acquired MR signals at coordinates $k$ and is labeled by the frame index $t$, and $N_t$ denotes the total number of acquired frames. A frame here refers to a cine-MR volume of a sufficient temporal resolution in the time series, so that the dynamic process under study can be considered quasi-static for each frame. In this study, we were interested in the respiration-induced motion, which is a major source of uncertainties in radiotherapy (Stemkens et al. 2018). Dynamic cine-MRI reconstruction aims to generate the moving sequence of the underlying subjects $\{z_t(x)\}_{t=0}^{N_t-1}$ in the image domain (i.e., time-varying cine-MR images), which are matched to the acquired signals in k-space $\{w_t(k)\}_{t=0}^{N_t-1}$ (Fessler 2010, Hansen et al. 2015). Here, $x$ denotes voxel coordinates of the reconstructed images. The reconstruction is formulated as an optimization problem with a regularization term:

$$\{\hat{z}_t\} = \underset{\{z_t\}}{\operatorname{argmin}} \left( \|F\{z_t(x)\} - \{w_t(k)\}\|^2 + \lambda\, R(\{z_t(x)\}) \right), \qquad (1)$$

where $F$ is an operator combining the coil sensitivity map and the Fourier transform matrix corresponding to the k-space sampling pattern. $R$ is the regularization term weighted by the factor $\lambda$. The data fidelity term (first term) of Eq. (1) enforces the data consistency between the reconstructed images $\{\hat{z}_t\}$ and the k-space MR acquisitions $\{w_t(k)\}$. The regularization term introduces prior knowledge of the images under study (i.e., sparsity in transformed domains) to facilitate the reconstruction and prevent overfitting in the optimization process.

To overcome the k-space undersampling issue, STINR-MR adopted a joint reconstruction and deformable registration-based approach, viewing each frame of the cine-MR images as a deformed version of a reference-frame image $z_{ref}(x)$:

$$z_t(x) = z_{ref}(x + d_t(x)), \qquad (2)$$

where $d_t(x)$ is the deformation vector field (DVF) at the cine frame *t*. Equation (2) assumes the existence of a reference frame and that the intra-scan motion can be described by these DVFs, which is supported by the fact that for images acquired within a single scan, the MR intensities are considered stable and the major variations are caused by anatomical motion. Note that the assumption of Eq. (2) excluded short-term physiological phenomena that may significantly change the MR intensities [e.g., contrast agents in dynamic contrast-enhanced MRI (Sourbron et al. 2013, Petralia et al. 2020)], which is not considered in this study. The reference-frame image $z_{ref}(x)$ serves as a template from which all cine-MR images are derived from, and itself may not necessarily correspond to an exact frame in the sequence $\{z_t\}$. Via Eqs. (1) and (2), STINR-MR decoupled the ill-posed spatiotemporal reconstruction problem into reconstructing a reference MR image $z_{ref}(x)$ and solving the corresponding dynamic motion $\{d_t(x)\}$, thus reducing the complexity of the reconstruction. In the following subsections, we first overviewed the workflow of STINR-MR, followed by details of the network architecture and training scheme. Afterwards, the dataset and evaluation schemes were presented.

## 2.2 STINR-MR workflow overview

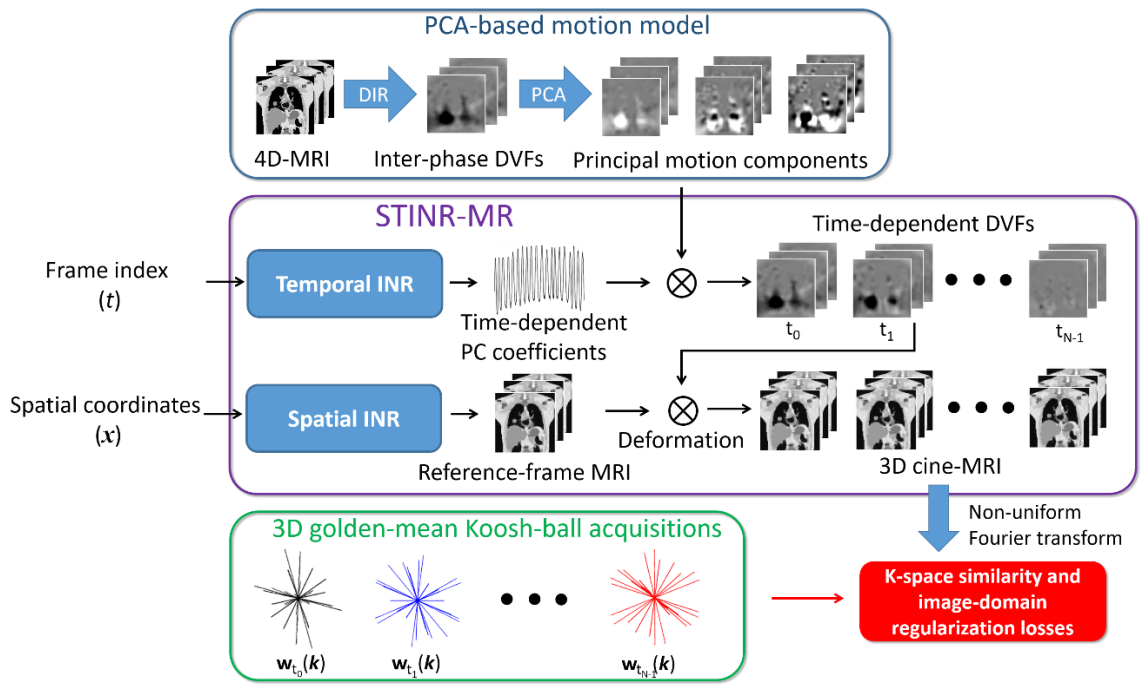

**Figure 1**. Overview of the workflow of 3D cine-MRI reconstruction (STINR-MR). The 3D cine-MRI reconstruction was based on a joint reconstruction and deformable image registration approach that each frame of the 3D cine-MRI was generated by deforming a reference-frame image (Eq. (2)). STINR-MR utilized a spatial implicit neural representation (INR) to reconstruct the reference-frame image, and a temporal INR to represent time-dependent motion. A principal component analysis (PCA)-based patient-specific motion model was incorporated into the framework to regularize the motion. The k-space readout was based on the 3D golden-mean Koosh-ball trajectory. The network training was driven by a k-space similarity loss and an image-domain regularization loss such that the k-space data of each reconstructed image match with the corresponding measured k-space signals. DIR: deformable image registration. PCA: principal component analysis. DVF: deformation vector field. INR: implicit neural representation.

Figure 1 illustrates the workflow of STINR-MR. STINR-MR consisted of a spatial INR and a temporal INR (middle box of Fig. 1). The spatial INR represents the reference-frame MR image and the temporal INR represents the intra-scan dynamic motion. Combining both INRs, 3D cine-MRI can be derived to represent spatiotemporal dynamics. In detail, the input into the spatial INR was a voxel coordinate $x$, and the output was the MR value at the queried coordinate (i.e., $z_{ref}(x)$). The entire volume of the reference frame can then be generated by querying all voxel coordinates within the region of interest. For the intra-scan motion (DVFs), the dimensionality of the solution space is extremely large, involving $\geq 10^8$ degrees of freedom (Huttinga et al. 2021). To regularize the solution of motion, we incorporated a principal component analysis (PCA)-based patient-specific motion model into the framework (top box of Fig. 1). PCA-based motion model introduced prior motion modes to significantly reduce the dimensionality of the unknown DVFs (Zhang et al. 2013, Zhang et al. 2019, Zhang et al. 2023). To derive the PCA-based motion model, a previously-acquired, motion-binned 4D-MRI can be used. Alternatively, the motion-binned 4D-MRI can also be directly derived from the cine-MRI acquisition. We obtained the inter-phase DVFs by registering the motion-binned 4D-MR images to the end-of-exhale bin, which is relatively stable with fewer artifacts (Vedam et al. 2001, Heerkens et al. 2014, Lever et al. 2014). The *principal motion components* can then be solved by performing PCA on the inter-phase DVFs of the 4D-MRI. The principal motion components can be considered as a basis set $\{e_i(x)\}$ spanning a Hilbert space and maximally accounting for the motion variance in the inter-phase DVFs $\{D_p(x)\}$:

$$\{e_i(x)\}_{i=1}^{N_{pc}} = \underset{\{e_i\}}{argmax}\left\{var\left[\sum_{p=1}^{N_{bin}} D_p(x) \cdot e_i(x)\right]\right\} \quad (3)$$

such that
$cov\left[\sum_{p=1}^{N_{bin}} D_p(x) \cdot e_i(x), \sum_{p=1}^{N_{bin}} D_p(x) \cdot e_j(x)\right] = 0$ for $i \neq j$ and $e_i(x) \cdot e_j(x) = \delta_{ij}$,

where $N_{pc}$ denotes the dimensionality of the space spanned by $\{e_i(x)\}$, $N_{bin}$ denotes the number of the motion bins of the 4D-MRI, and *var* and *cov* respectively denote the variance and covariance of their arguments. $\delta_{ij}$ denotes the Kronecker delta, and the inner product is defined in the Hilbert space of the motion fields. An arbitrary respiratory DVF can be represented as a linear combination of these principal motion components. Here, we used the first three principal motion components (together with the mean inter-phase DVF) as the basis, as the first three components were shown sufficient to accurately describe the respiratory motion (Li et al. 2011). Through this strategy, the PCA-based motion model reduced the dimensionality of the unknown DVFs from $\geq 10^8$ to 9. With the PCA-based motion model, we used a temporal INR to represent the PCA weightings, in the form of nine PC coefficients (i.e., three principal motion components × three Cartesian directions) at each queried frame index. The principle components, scaled by the weightings output from the temporal INR, were superposed to generate frame-specific DVFs:

$$d_t(x) = e_0(x) + \sum_{i=1}^{3} w_i(t) \times e_i(x), \quad (4)$$

where $e_0(x)$ is the mean DVF of $\{D_p(x)\}$, $e_i(x)$ is the $i^{th}$ principal motion component, and $w_i(t)$ is the corresponding PC weighting. The time sequences of the PC weightings, output via the temporal INR, form the time series of the motion fields $\{d_t(x)\}$ to capture the dynamic motion. Finally, the 3D cine-MR images were reconstructed by applying the sequence of $\{d_t(x)\}$ to the reference-frame MRI, as in Eq. (2).

As shown in the workflow, the spatial INR and the temporal INR were jointly solved (trained), by matching the projected k-space data of reconstructed 3D cine-MR images to the acquired k-space data. The training was purely driven by the acquired data of each dynamic MR acquisition in an iterative fashion, thus allowing 'one-shot' learning. In addition to the k-space data fidelity loss, we also introduced an image domain regularization term, the total variation of the reference-frame MR image (Rudin et al. 1992), to regularize the reconstruction quality and improve the convergence speed. The joint training scheme allowed concurrent update and refinement of the reference frame and the intra-scan motion via all k-space data, thus improving the overall accuracy and consistency throughout the time series.

2.3 Network architectures and the training scheme

*2.3.1 The spatial implicit neural representation*

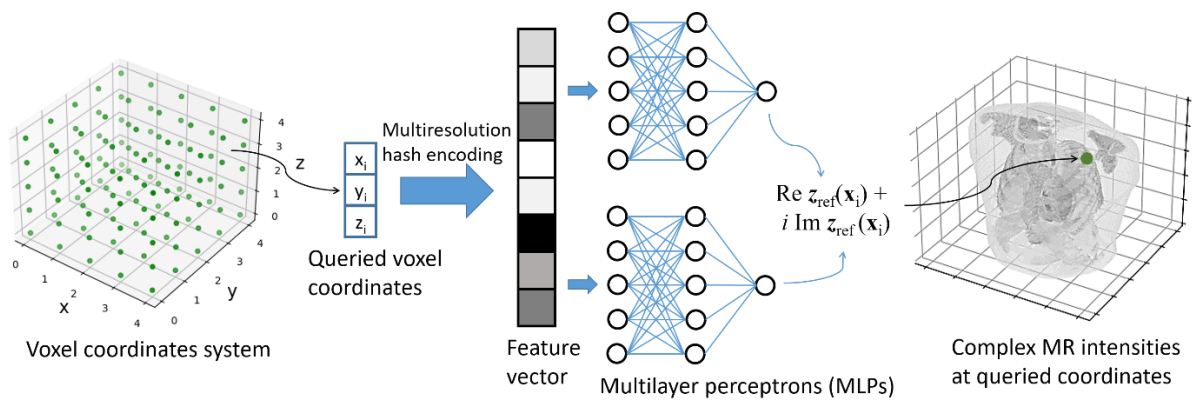

**Figure 2**. Workflow and network architecture of the spatial INR. The spatial INR network took a voxel coordinate $x$ as the input, and output the complex-valued MR intensity at the queried voxel. The input coordinate was first mapped to a higher dimension space by a multiresolution hash encoding scheme, and the resulting feature vector was input into a subsequent structure of multilayer perceptrons (MLPs). Two independent MLPs were respectively used to represent the real and imaginary parts of the image. The volume reconstruction was achieved by querying all voxel coordinates within the region of interest.

Figure 2 illustrates the workflow of the reference-frame MRI reconstruction. As mentioned in Sec. 2.2, the spatial INR mapped 3D voxel coordinates to the corresponding complex-valued MR intensities. The INR was constructed via multilayer perceptrons (MLPs) to serve as a universal function approximator. As the MLPs have shown difficulties in learning high-frequency image features directly (Tancik et al. 2020), the input coordinates need to be pre-processed by a learning-based position encoding scheme before inputting into the MLPs, to promote the learning of high-frequency features. We used the multiresolution hash encoding (Muller et al. 2022), which mapped the 3D space to a higher dimension space, using a spatial hash function and a multiresolution hierarchy of hash tables (Fig. 3). The hash tables were learning-based with trainable parameters, allowing efficient and adaptive encoding. The output of the hash encoding was a feature vector whose length depended on the number of multiresolution levels. The multiresolution hash encoding has shown advantages over other encoding schemes in terms of the representation quality, the versatility of usage, and the training speed (Muller et al. 2022). In addition, by the multiresolution hash encoding, the depth of the MLPs can be reduced, allowing smaller and more efficient architectures to be deployed. Therefore, the training time can be significantly shortened. For the multiresolution hash encoding, we used hyper-parameter values recommended by the literature (Muller et al. 2022), and they were summarized in Table 1. The range of the voxel coordinate system was scaled between -1 and 1 prior to the hash encoding.

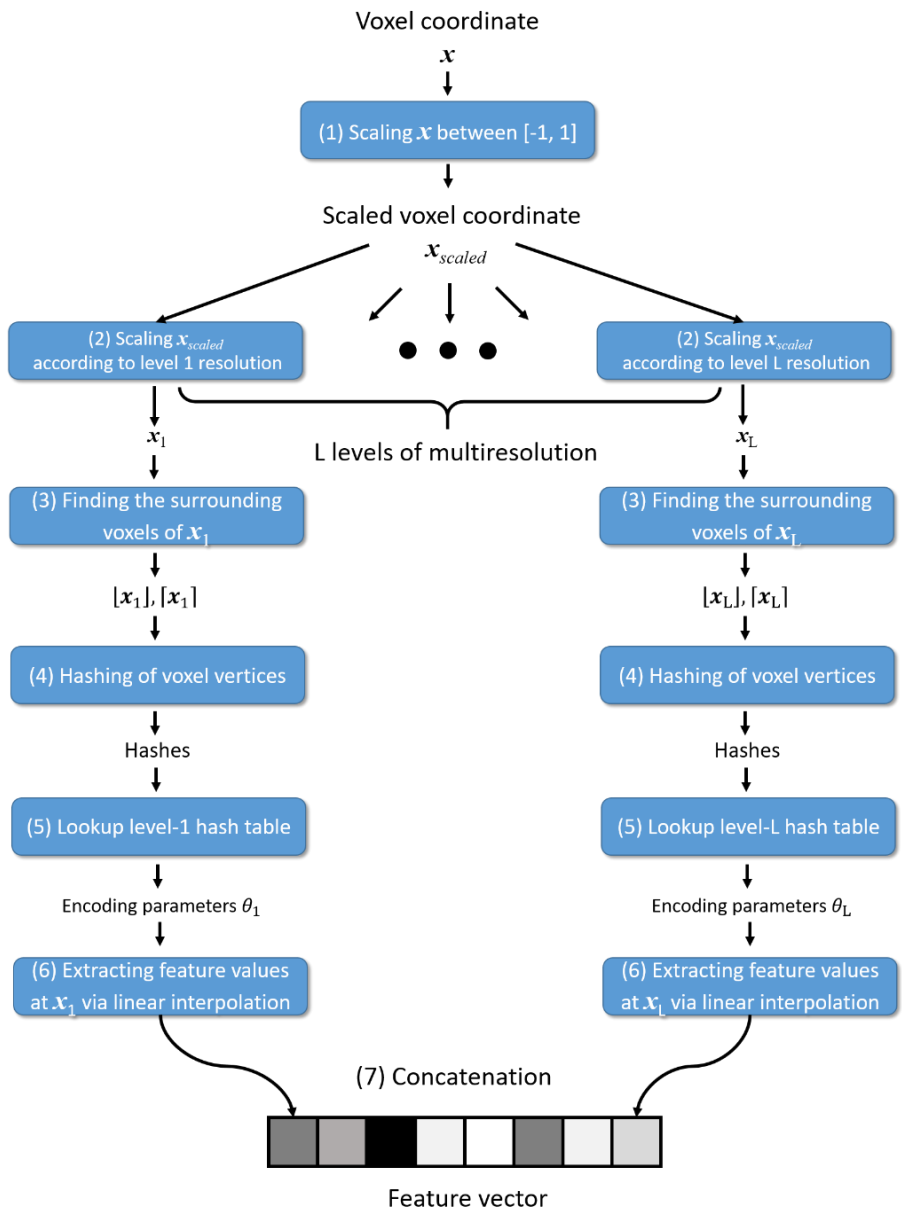

**Figure 3.** Workflow of the multiresolution hash encoding (Muller et al. 2022). The multiresolution hash encoding mapped a voxel coordinate to a feature vector in a higher dimension space via learnable hash tables. The encoding scheme was through a multiresolution approach that progressively increases the spatial resolution at higher levels. At each level, the encoding scheme sets up a grid of vertices with integer indexes based on the resolution of the level. Then the encoding scheme first mapped the input voxel coordinate $x$ to this grid system by scaling the coordinate $x$ in steps (1) and (2). Through steps (3) to (5), a hash function mapped the indexes of the surrounding vertices of the scaled coordinate to the learnable hash table to retrieve the encoding parameters. The feature values of the voxel coordinate were subsequently extracted based on the relative position of the voxel to its surrounding vertices in step (6), via linear interpolations of the encoding parameters. Finally, the extracted feature values of all levels were concatenated in step (7).

**Table 1**. Hyper-parameters of the multiresolution hash encoding.

| Hyper-parameter | Value |
|---|---|
| Number of levels | 16 |
| Maximum entries per level | $2^{19}$ |
| Number of feature dimensions per entry | 2 |
| Coarsest resolution | 16 |
| Finest resolution | 10,509 |

Since MR images are complex-valued, two independent MLPs were used for the spatial INR to represent the real and imaginary parts of the image, respectively. Each MLP comprised an input, a hidden, and an output layer, whose feature numbers were 32, 32, and 1, respectively. Both MLPs shared the same hash-encoded feature vector. We used the same periodic activation function as a previous study (Sitzmann et al. 2020), and initialized learnable parameters of the MLPs in a similar way.

*2.3.2 The temporal implicit neural representation*

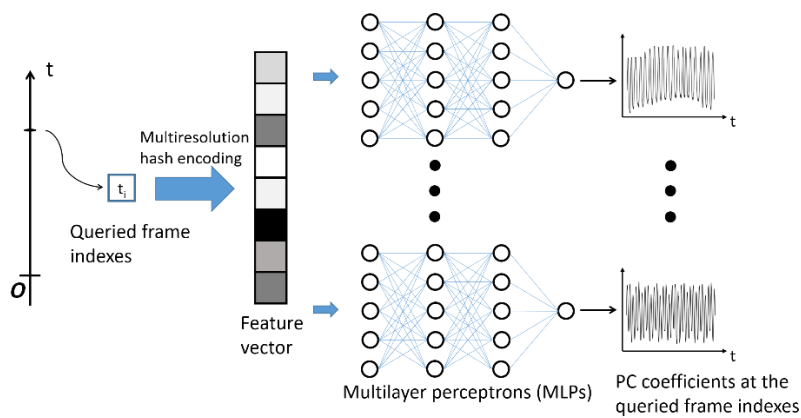

**Figure 4**. Workflow and network architecture of the temporal INR. The temporal INR took as input a frame index of the MR acquisition and output principal component (PC) weightings/coefficients at the queried frame index. Similar to the spatial INR (Fig. 2), the input coordinates were first mapped to a higher dimension space by the multiresolution hash encoding. Nine parallel MLPs were used to map the encoded feature vector to nine frame-dependent PC coefficients. The complete temporal sequence of PC coefficients were obtained by querying all frame indexes within the sequence.

The temporal INR network represents the intra-scan motion (Fig. 4). The input was a frame index, and the output were nine PC weightings/coefficients at the queried frame to compose the frame-specific DVFs. The temporal INR shared a similar network architecture as the spatial INR, consisting of the multiresolution hash encoding and nine parallel MLPs. Each MLP had one input layer, one output layer, and two hidden layers with rectified linear unit activation functions. The same set of hyper-parameters of the spatial INR (Table 1) was used in the temporal INR. The feature numbers of the input and hidden layers were 32, and the feature number of the output layer was 1. Before feeding into the hash encoding, the frame index was scaled between -1 and 1. In addition, the scaled frame indexes were randomly perturbed within their frame intervals (i.e., the temporal resolution of the cine-MRI) with a Gaussian noise to force the temporal INR to learn continuous representations of PC coefficients as a function of the scaled frame index (Reed et al. 2021).

*2.3.3 The progressive training scheme*

Owing to the interplay between the reference-frame reconstruction (spatial INR training), the motion solution (temporal INR training), and the limited k-space data in each frame, training the spatial and temporal INRs simultaneously from scratch was found slow and challenging. To address this challenge, we initialized the spatial INR prior to the joint training (i.e., a warm start for the joint training), and designed a STINR-MR training scheme with progressively added complexity to avoid the local minimum (Zhang et al. 2023). The training scheme contained three stages. In the first two stages, only the spatial INR was trained (without the motion solution) to construct an approximated reference frame, and the joint spatial and temporal INR training was introduced later at stage 3. To generate an initial reference-frame image for the warm start, the MR acquisitions $\{w_t(k)\}_{t=0}^{N_t-1}$ were first sorted into 10 respiratory phases, and the subset corresponding to the end-of-exhale phase was selected to reconstruct an approximated reference frame by non-uniform fast Fourier transform (NUFFT) (Muckley et al. 2020). The phase sorting can be based on an external signal (e.g., optical surface imaging (Bertholet et al. 2019, Padilla et al. 2019) or pressure belt (Li et al. 2006)) or self-navigating signals directly extracted from the k-space data (Larson et al. 2005, Brau et al. 2006, Grimm et al. 2015). For stage 1, the spatial INR was learned directly from the NUFFT-reconstructed reference-frame MR image, by minimizing the L2 similarity loss in the image domain. In stage 2, the similarity loss was instead evaluated and minimized in k-space using the raw data from the end-of-exhale phase, which helps to address the undersampling artifacts from the NUFFT reconstruction (see Fig. S-1 in Supplementary materials). For Stage 3, both the spatial INR and the temporal INR were activated for learning, as shown in Fig. 1.

2.4 Data curation and evaluation schemes

STINR-MR was evaluated using the XCAT digital phantom (Segars et al. 2010) and a public-available dataset of a healthy human subject (Huttinga et al. 2021). The XCAT phantom can simulate various respiratory motion with 'ground-truth' images to allow quantitative evaluations and analyses. Therefore, the XCAT study served to demonstrate the feasibility and accuracy of the proposed framework as a proof-of-concept. The human subject study served to further demonstrate the applicability of the STINR-MR on real-world data. Because of the distinct nature of the two datasets, we separately described them in the following subsections.

*2.4.1 The XCAT phantom study*

We simulated 3D cine-MR images of XCAT using different respiratory motion to evaluate STINR-MR. To simulate the cine-MR images, we first used XCAT to generate a 4D-MRI set of 10 respiratory phases (with a 5-s cycle and 20-mm diaphragm peak-to-peak motion), computed inter-phase DVFs (relative to the end-of-exhale phase) via Elastix (Klein et al. 2010), and derived principal motion components of the inter-phase DVFs via PCA. Different motion scenarios were then simulated by rescaling the principal motion components to generate intra-scan DVFs according to different motion curves and then using these DVFs to deform the end-of-exhale MR volume to 3D cine-MRI series. The end-of-exhale XCAT MR volume covered the whole thorax and the upper portion of the abdomen. A spherical lung tumor of 30-mm diameter was inserted into the lower lobe of the right lung, serving as a target for assessing the accuracy of solved motion. The volume size was 100×100×100 with an isotropic 4-mm resolution. Since the XCAT phantom generated magnitude-only MR images, complex-valued images were simulated by adding spatial phase modulation to the real-valued, end-of-exhale MR image volume. The spatial phase modulation was simulated as a superposition of four sinusoidal oscillations (Zhu et al. 2018, Terpstra et al. 2020). The wave number of each sinusoid was randomly selected between [0.0033 mm$^{-1}$, 0.02 mm$^{-1}$] with a random orientation and a random phase shift. After superposing the four sinusoidal oscillations, the amplitude was normalized between 0 and 2π. The normalized phase map was used as the exponent to generate complex-valued phase modulation. For each simulated motion scenario, we applied the same phase modulation to the end-of-exhale MR image volume, and deformed the volume to complex-valued 3D cine-MR images via the simulated, scenario-specific intra-scan DVFs.

Specifically, STINR-MR was evaluated for different motion/anatomical scenarios including: (i) various types of regular/irregular respiratory motion; and (ii) inter-scan anatomical variations between the original 4D-MRI and the cine MR scan. For (i), six types of respiration motion with various degrees of complexity were simulated. Table 2 highlights the characteristics of the motion scenarios, and Fig. 5 shows the corresponding center-of-mass motion trajectories of the lung tumor along the superior-inferior (SI) direction. All motion trajectories correspond to a 180-s scan and 1,826 ($N_t$) cine frames (each frame having a temporal resolution of 98.6 ms). Specifically, S1 was the simplest motion scenario, having small variations of the breathing amplitude along a constant baseline. On the basis of S1, S2 added a 7-mm downward baseline shift at around 90 s into the scan. S3 contained both amplitude variations and baseline shifts. S4 had a change of the breathing period and the amplitude starting from 60 s into the scan. S5 included a slow breathing motion with gradually decreasing motion amplitudes. S6 was the most complex scenario involving combined variations of breathing period, amplitude, and baseline. For (ii), we simulated inter-scan anatomical variations by reducing the lung tumor size of the end-of-exhale MR volume (from 30 mm to 15 mm in diameter), before mapping it to 3D cine-MR images using the intra-scan DVFs of the motion scenario S1.

**Table 2**. Summary of motion characteristics of six motion scenarios in the XCAT phantom study.

| Motion scenario | Motion characteristics |
| --- | --- |
| S1 | Amplitude variations |
| S2 | Baseline shifts |
| S3 | Amplitude variations and baseline shifts |
| S4 | Respiratory period/amplitude variations |
| S5 | Slow breathing with a small amplitude variation |
| S6 | Combination of respiratory period/amplitude variations and baseline shifts |

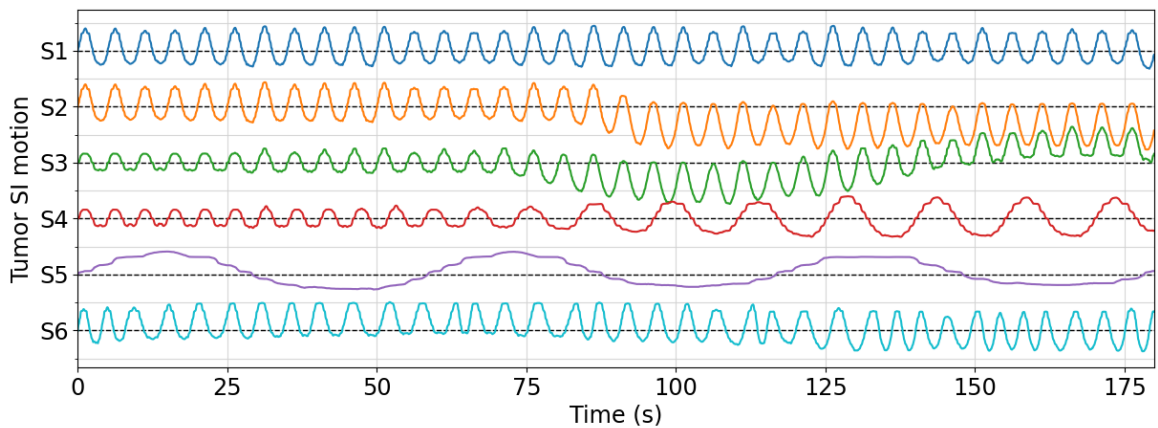

**Figure 5**. Lung tumor center-of-mass trajectories in the superior-inferior (SI) direction. See Table 2 for the descriptions of the motion scenarios (S1-S6).

From the simulated 'ground-truth' complex-valued 3D cine-MR images, we generated the k-space data assuming free-breathing MR acquisitions, for STINR-MR reconstruction and evaluation. For simplicity, we considered the MR acquisitions involved a single coil with a uniform sensitivity map covering the whole field-of-view. We used the gradient echo-based pulse sequences, with the k-space data acquisition simulated via 3D golden-mean Koosh-ball trajectories (Winkelmann et al. 2007, Chan et al. 2009, Feng 2022). The Koosh-ball trajectory was non-Cartesian and comprised of readout lines in the radial directions (i.e., spokes). Each readout line passed through the k-space origin, with its orientation order following the golden-mean algorithm (Chan et al. 2009). Via the Koosh-ball trajectory, the central region of k-space was oversampled to be more motion robust. The data can be easily sorted by motion for self-navigation, which renders the Koosh-ball trajectory particularly suitable for dynamic 3D-cine MRI (Lingala et al. 2016, Stemkens et al. 2018). In addition, the gold-mean angular sampling scheme minimized the coherent interference of undersampling artifacts. Although the Koosh-ball trajectory was the focus of this study, STINR-MR can be readily applied to other 3D trajectories (e.g., (Liao et al. 1997, Burdumy et al. 2017)), as the image reconstruction and motion solution were irrelevant to the k-space trajectory specifics.

For the k-space simulation, we used a TR = 5.8 ms (Deng et al. 2016), corresponding to 17 spokes per frame (each frame has a 98.6 ms temporal resolution). In our evaluation, we also tested using even fewer spokes per frame to reconstruct more frames and further increase the temporal resolution. In detail, we used 4, 8, or 17 spokes to represent a frame for STINR-MR reconstruction, which corresponded to 23.2 ms, 46.4 ms, and 98.6 ms in temporal resolution, for the S1 motion scenario study.

STINR-MR required a PCA-based motion model as input (Fig. 1), which can come from two sources: (1). the PCA model from the originally-simulated 4D-MRI (as described above), which in clinical practice can be a previously-acquired 4D-MRI that offers offline information (offline PCA); and (2). the PCA model directly derived from 4D-MRIs reconstructed using the online cine MR acquisitions (online PCA). In cases that a previously-acquired 4D-MRI may not be available, we can sort the k-space data of cine MR acquisitions into 10 phases to reconstruct an online 4D-MRI via NUFFT, and perform PCA without relying on any prior data. In this study, we evaluated both approaches and compared their accuracy.

*2.4.2 Evaluation metrics of the XCAT study*

We evaluated the accuracy of the reconstructed 3D cine-MR images and the accuracy of the tumor motion solved by intra-scan DVFs, by comparing them with the simulated 'ground-truth'. The reconstructed reference-frame MR images were visually examined, and the whole sequence of 3D cine-MR images were quantitatively evaluated using the relative error (RE) metric:

$$\text{RE} = \sqrt{\frac{\sum_x |\hat{z}_t(x) - z_t^{gt}(x)|^2}{\sum_x |z_t^{gt}(x)|^2}}, \qquad (5)$$

where $z_t^{gt}$ denotes the 'ground-truth' image. The accuracy of the tracked tumor motion by solved intra-scan DVFs was evaluated using the tumor center-of-mass error (COME) and the Dice similarity coefficient (DSC). The COME measures the center-of-mass distance between the DVF-propagated tumor location and the 'ground-truth' tumor location. The DSC is defined by

$$\text{DSC} = \frac{2 \times |Y \cap Y^{gt}|}{|Y| + |Y^{gt}|}, \qquad (6)$$

where $Y$ and $Y^{gt}$ denote the DVF-propagated and the 'ground-truth' tumor contours, respectively.

*2.4.3 The human subject study*

In addition to the XCAT simulation study, we also evaluated STINR-MR using a free-breathing scan of a healthy human subject acquired by a 1.5-T MRI scanner (Ingenia, Philips Healthcare) (Huttinga et al. 2021). For the k-space acquisition, the phase array consisted of 12 anterior and 12 posterior receive coils, and the sensitivity map and the noise covariance matrix were provided for each coil. The pulse sequence was a steady-state spoiled gradient echo sequence. The TR and echo time were 4.4 ms and 1.8 ms, respectively, and the flip angle was 20°. Same as the XCAT simulation study, the k-space was acquired via a 3D golden-mean Koosh-ball radial trajectory. The total scan duration was 297.4 s, resulting in 67,280 radial spokes with 232 readout points per spoke. The first 900 spokes were discarded to allow the system to reach a steady state. The scan covered the thoracic and abdominal regions.

Different from the XCAT simulation study, for the human subject study there is no prior 4D-MRI available to build the PCA motion model. Thus, we built an online PCA model directly using the available k-space data. In detail, we extracted a surrogate signal representing the respiratory motion from the k-space signals (Huttinga et al. 2022). The k-space signals of each coil at the origin $\{w_t(k = 0)\}$ were extracted from all sequential radial readouts, and consolidated as a 24-channel time series. It was subsequently processed by a low-pass filter using the Kaiser window method (Kaiser et al. 1980) to remove high-frequency noises. PCA was then performed on the filtered time series, and the principal component with the largest spectral density in the frequency range between 0.1 Hz and 0.5 Hz (corresponding to the respiratory motion frequency range) was selected as the surrogate signal. Based on the surrogate signal, the radial spokes were sorted into 10 respiratory phases and reconstructed into a 4D-MRI using NUFFT (Muckley et al. 2020). The reconstructed image size was 150×150×150, with a 3.0×3.0×3.0 mm$^3$ resolution. From the 4D-MRI, a PCA-based motion model was generated, as described in Sec. 2.2.

For the human subject study, the L2 similarity loss of STINR-MR was defined for each of the 24 coils and then summed together. To achieve a balance between the noise suppression and the temporal resolution, we binned 68 radial spokes per coil into a frame, which corresponds to a temporal resolution of 299.2 ms. Since no 'ground-truth' was available for the human study, STINR-MR's performance was assessed by visual inspection and quality evaluation of the reconstructed reference-frame MR image. For quality evaluation, we measured the sharpness of the reconstructed reference frame using gradient- and variance-based metrics (Ferzli et al. 2005). The gradient metric is defined as

$$\text{gradient} = \frac{1}{N_{voxel}} \sum_{x} \left| \nabla \left| z_{ref}(x) \right| \right|, \qquad (7)$$

where $N_{voxel}$ is the number of voxels of the reference-frame MR image. The variance metric was calculated as the mean variance of the whole reference-frame MR image (Ferzli et al. 2005). For both metrics, higher values indicate sharper images with less motion blurriness. We also compared the liver center-of-mass motion tracked by STINR-MR with the k-space surrogate's motion. The liver center-of-mass was calculated by contouring the liver in the reference frame and then propagating the contour by the intra-scan DVFs solved by STINR-MR.

*2.4.4 Other hyper-parameters of the STINR-MR framework and the training details*

The Adam optimizer was used for STINR-MR training. Under the progressive training scheme (Sec. 2.3.3), the learning rate of the MLPs in the spatial INR were reset at the beginning of each stage. For the XCAT study, we used learning rates of 1×10$^{-3}$, 2×10$^{-5}$, and 2×10$^{-6}$ empirically for the first, second, and the last stages, respectively. For the human subject study, we used learning rates of 1×10$^{-3}$, 1×10$^{-10}$, and 1×10$^{-12}$ for the three stages, respectively. For the XCAT study, the first, second, and the last stages were trained by 500, 1500, and 1000 epochs, respectively. For the human subject study, the corresponding epochs were 500, 300, and 1100, respectively. For the joint training of the third stage, one epoch contained 60 frames randomly selected from the MR acquisitions, which was determined to balance the training speed and to avoid the temporal aliasing while being bounded by the available memory in the graphic processing unit (NVIDIA A100). The weighting factors λ of the XCAT and the human subject studies were empirically set to $2\times10^{-4}$ and $2\times10^{-6}$, respectively. The overall training time was ~20 minutes and ~100 minutes for the XCAT and the human subject studies, respectively. The training time difference was mainly due to the size differences of the reference frame (XCAT: $100\times100\times100$ with a 4-mm isotropic resolution; Human subject: $150\times150\times150$ with a 3-mm isotropic resolution), the k-space complexity (single-channel vs. multi-channel), and the underlying complexity of the reconstructed anatomy.

*2.4.5 The comparison study with MR-MOTUS*

STINR-MR was compared with MR-MOTUS (Huttinga et al. 2020, Huttinga et al. 2021), a model-based and non-rigid motion estimation method that was recently developed. MR-MOTUS had three features distinct from STINR-MR: (i) the model was formulated for k-space data of a single channel, so multi-coil data had to be compressed into a single virtual channel prior to the MR-MOTUS reconstruction. The coil compression was through a linear superposition of the sensitivity maps where the superposition coefficients were determined such that the compressed sensitivity map maximizes the spatial homogeneity. (ii) The reference-frame MR image was from a previously-acquired MRI, or reconstructed from the motion-sorted k-space data, with no additional refinement during the motion estimation stage. (iii) MR-MOTUS used a low-rank motion model to regularize the motion estimation. The low-rank motion model assumed the time-dependent motion fields to be partially separable (Zhao et al. 2012), i.e., the motion field can be written as a summation of products of a spatial function and a temporal function, similar to our PCA model. The spatial functions represented motion modes, and the temporal functions mapped the time-varying weightings of these motion modes. Both spatial and temporal functions were parametrized by B-splines for dimension reduction.

To meet the computational and memory demands of the low-rank MR-MOTUS algorithm, we downsampled the reference-frame MR images of the XCAT and the human subject studies to resolutions of $8\times8\times8$ mm$^3$ and $6.7\times6.7\times6.7$ mm$^3$, respectively (Huttinga et al. 2021). In addition, the whole MR data sequences were partitioned into smaller batches, and the motion fields were estimated for each batch separately and independently to meet the memory limit. The XCAT and the human subject studies used three and eight batches, respectively. To achieve a reasonable computation time (~62 minutes/batch for the XCAT study), the temporal resolution for the XCAT study was set to 197.2 ms (34 spokes), with the number of motion modes set to 2. The temporal resolution for the human subject study was 299.2 ms (same as STINR-MR) with the number of motion modes set to 3, which takes ~45 minutes/batch to compute.

## 3. Results

3.1 The XCAT phantom study

*3.1.1 Reconstruction accuracy under different respiratory motion variations*

Figure 6 visually compares reconstructed reference-frame MR images by three methods (STINR-MR: offline PCA; STINR-MR: online PCA; and MR-MOTUS). STINR-MR with offline PCA presented images with the highest quality for all motion scenarios, while the reference-frame images of MR-MOTUS showed strip artifacts due to undersampling and motion. STINR-MR with online PCA presented images with overall good quality, while some artifacts can be observed due to the inaccuracy of the online-derived PCA models (due to irregular and non-period motion, intra-phase motion, and sorting errors). Table 3 summarizes the mean relative error metric averaged over the entire 3D cine-MR image series.

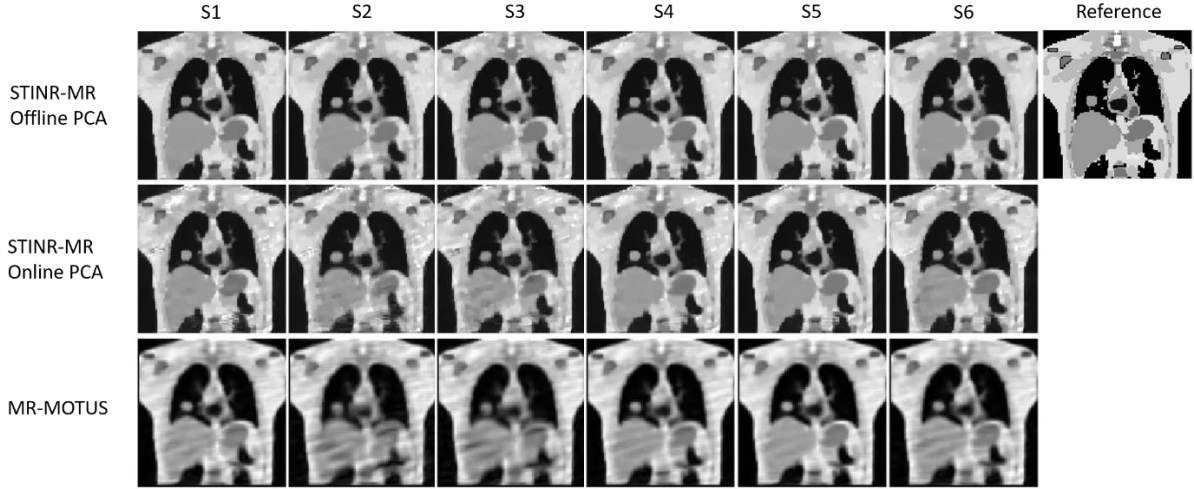

**Figure 6**. Comparison of reconstructed reference-frame MR images for the six motion scenarios (S1-S6) using three methods. Each column (columns 1-6) shows the reconstructed reference-frame images of one motion scenario, and column 7 shows the reference image of the XCAT simulation.

**Table 3**. Mean±S.D. relative error of reconstructed 3D cine-MR images over the whole motion sequence for the various motion scenario study.

| Motion scenario | STINR-MR Offline PCA | STINR-MR Online PCA | MR-MOTUS |
|---|---|---|---|
| S1 | 0.178±0.003 | 0.233±0.006 | 0.279±0.007 |
| S2 | 0.187±0.005 | 0.258±0.009 | 0.304±0.009 |
| S3 | 0.188±0.004 | 0.247±0.015 | 0.296±0.006 |
| S4 | 0.187±0.003 | 0.218±0.006 | 0.274±0.005 |
| S5 | 0.184±0.003 | 0.218±0.006 | 0.273±0.008 |
| S6 | 0.186±0.003 | 0.230±0.005 | 0.280±0.008 |

**Table 4**. Lung tumor localization accuracy for the six motion scenarios, measured by the tumor center-of-mass error (COME) and the Dice similarity score (DSC). The values were presented in terms of mean and standard deviation.

| Motion scenario | COME (mm) | | | DSC | | |
|---|---|---|---|---|---|---|
| | STINR-MR Offline PCA | STINR-MR Online PCA | MR-MOTUS | STINR-MR Offline PCA | STINR-MR Online PCA | MR-MOTUS |
| S1 | 0.9±0.4 | 1.4±0.7 | 3.5±0.9 | 0.92±0.02 | 0.90±0.03 | 0.80±0.04 |
| S2 | 0.9±0.4 | 2.7±1.2 | 3.1±1.0 | 0.89±0.02 | 0.83±0.05 | 0.73±0.05 |
| S3 | 1.1±0.5 | 2.1±1.4 | 3.2±1.0 | 0.91±0.02 | 0.85±0.05 | 0.78±0.04 |
| S4 | 1.0±0.4 | 1.3±0.5 | 3.2±1.0 | 0.91±0.02 | 0.89±0.02 | 0.81±0.05 |
| S5 | 0.9±0.5 | 1.3±0.6 | 3.5±1.0 | 0.92±0.02 | 0.89±0.02 | 0.80±0.05 |
| S6 | 1.2±0.4 | 1.4±0.8 | 4.0±1.1 | 0.92±0.03 | 0.89±0.03 | 0.78±0.05 |

Table 4 summarizes the lung tumor localization accuracy measured over the whole sequences of 3D cine-MR images. Both variants of STINR-MR outperformed MR-MOTUS and achieved sub-voxel localization accuracy. A comparison of the tumor center-of-mass motion in the SI direction as a function of time was given in Supplementary materials due to the limitation of space (Fig. S-2).

*3.1.2 Reconstruction accuracy under inter-scan tumor size variations*

Figure 7 presents the reconstructed reference-frame MR images of the three methods in the case of tumor shrinkage, and Table 5 summarizes the mean and standard deviation of the relative errors averaged over the whole sequence of the 3D cine-MRI.

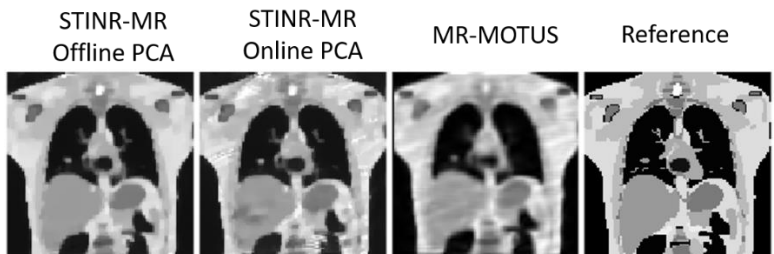

**Figure 7**. Comparison of the reconstructed reference-frame MR images under the scenario of inter-scan tumor size shrinkage for the three methods (based on motion scenario S1). The rightmost panel shows the reference image of the XCAT simulation.

**Table 5**. Mean±S.D. relative error of reconstructed 3D cine-MR images over the whole motion sequence for the inter-scan tumor shrinkage study.

| Tumor diameter (mm) | STINR-MR Offline PCA | STINR-MR Online PCA | MR-MOTUS |
|---|---|---|---|
| 15 | 0.193±0.006 | 0.236±0.005 | 0.277±0.007 |

**Table 6**. Lung tumor localization accuracy for the inter-scan tumor shrinkage study, measured by the tumor center-of-mass error (COME) and the Dice similarity score (DSC). The values were presented in terms of mean and standard deviation.

| Tumor diameter (mm) | COME (mm) | | | DSC | | |
|---|---|---|---|---|---|---|
| | STINR-MR Offline PCA | STINR-MR Online PCA | MR-MOTUS | STINR-MR Offline PCA | STINR-MR Online PCA | MR-MOTUS |
| 15 | 2.2±0.9 | 1.6±1.0 | 4.8±1.2 | 0.78±0.07 | 0.79±0.05 | 0.61±0.08 |

Table 6 summarizes the tumor localization accuracy for the inter-scan tumor shrinkage scenario. Compared with the 30-mm tumor case (S1 scenario in Table 4), the offline-PCA STINR-MR variant saw a larger increase in COME than that of the online-PCA variant, potentially due to the inter-scan anatomical change that affects the accuracy of the offline PCA motion model. The DSC values also saw a decrease compared to Table 4, which is a combined effect of the reconstruction accuracy and the increased sensitivity of DSC to smaller tumor sizes.

*3.1.3 Reconstruction accuracy under temporal resolution variations*

Figure 8 compares the reference-frame MR images reconstructed under different temporal resolutions for the offline-PCA and online-PCA variants, and Table 7 presents the relative errors averaged over the temporal sequences. MR-MOTUS was not tested for higher temporal resolutions due to hardware (memory) constraints. Both variants of the STINR-MR technique showed robustness to the temporal resolution variations (corresponding to varying numbers of radial spokes assigned to each temporal frame).

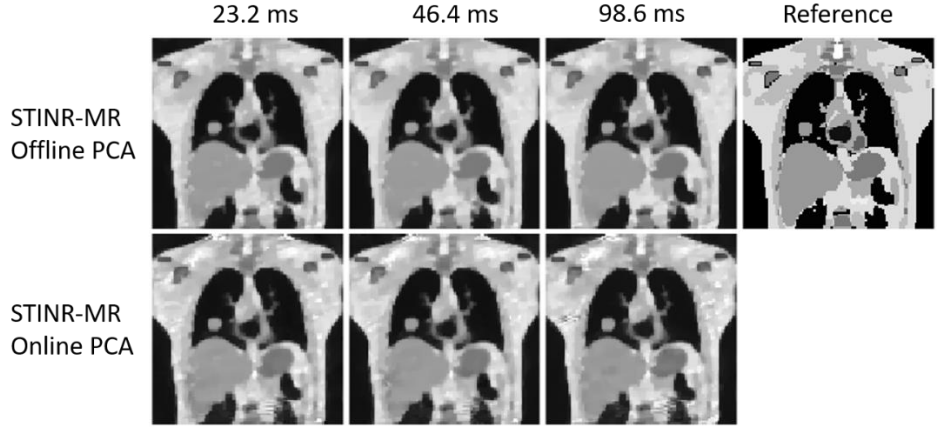

**Figure 8**. Comparison of reference-frame MR images reconstructed under various temporal resolutions of STINR-MR reconstruction (based on motion scenario S1). The rightmost panel shows the reference image of the XCAT simulation.

**Table 7**. Mean±S.D. relative error of reconstructed 3D cine-MR images over the whole motion sequence for the temporal resolution variation study.

| Temporal resolution (ms) | STINR-MR Offline PCA | STINR-MR Online PCA |
|---|---|---|
| 23.2 | 0.184±0.003 | 0.233±0.006 |
| 46.4 | 0.181±0.003 | 0.232±0.006 |
| 98.6 | 0.178±0.003 | 0.233±0.006 |

**Table 8**. Lung tumor localization accuracy for the temporal resolution variation study, measured by the tumor center-of-mass error (COME) and the Dice similarity score (DSC). The values were presented in terms of mean and standard deviation.

| Temporal resolution (ms) | COME (mm) | | DSC | |
|---|---|---|---|---|
| | STINR-MR Offline PCA | STINR-MR Online PCA | STINR-MR Offline PCA | STINR-MR Online PCA |
| 23.2 | 1.0±0.5 | 1.3±0.6 | 0.92±0.02 | 0.89±0.03 |
| 46.4 | 1.1±0.5 | 1.3±0.7 | 0.92±0.02 | 0.91±0.03 |
| 98.6 | 0.9±0.4 | 1.4±0.7 | 0.92±0.02 | 0.90±0.03 |

Table 8 summarizes the tumor COME and DSC under the three temporal resolutions. Both variants of STINR-MR were insensitive to the temporal resolution variations and achieved sub-voxel localization accuracy.

3.2 The human subject study

Figure 9 compares the reference-frame MR images reconstructed by STINR-MR (via online PCA) and by MR-MOTUS. A line profile across the liver was also compared. The STINR-MR reference frame showed sharper organ boundaries, while the MR-MOTUS reference frame appeared overly-smoothed with blurriness potentially caused by intra-scan motion. The MR intensities of STINR-MR remained at a higher level outside of the subject due to the zero-valued sensitivity maps at the field-of-view edges. The calculated gradient metrics for STINR-MR and MR-MOTUS were 0.0151 and 0.0125, respectively. The corresponding variance metrics were 0.0079 and 0.0073, respectively. Both metrics showed that STINR-MR provided sharper reference-frame MR images after joint reconstruction and deformable registration, which yielded more anatomical details to guide the diagnosis and/or treatments.

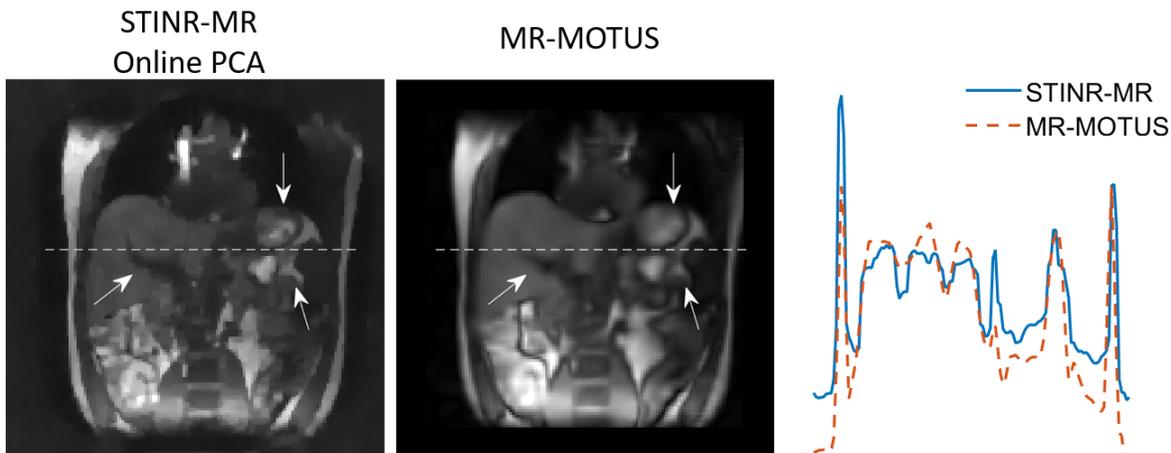

**Figure 9**. Comparison of reconstructed reference-frame MR images and the corresponding line profiles for the human subject study. The corresponding location of the line profiles was indicated by the horizontal dashed lines, and the arrows highlighted the over-smoothed regions of MR-MOTUS.

Figure 10 compares the tracked liver center-of-mass motion in the SI, anterior-posterior (AP), and left-right (LR) directions, by STINR-MR and MR-MOTUS. For comparison, the surrogate signal directly extracted from the k-space was plotted in Fig. 10a (Sec. 2.4.3). The surrogate signal was extracted from the origin of the k-space data via denoising and PCA to show the general motion trend, and may not fully represent the detailed liver motion. Along the SI direction, the amplitude variations of the surrogate signal at around 190 s and the breathing period variations at around 215 s were reproduced by both STINR-MR and MR-MOTUS. Overall, STINR-MR and MR-MOTUS solved similar motion curves in the SI direction, but STINR-MR had slightly larger SI motion amplitudes than MR-MOTUS. In addition, MR-MOTUS solved smaller AP motion amplitudes than STINR-MR, and it also had a relative baseline shift in the LR direction. The smaller motion amplitudes of MR-MOTUS could be due to the motion blurriness and over-smoothing observed in its reference-frame MR image (Fig. 9). Considering that the spatial resolution of the MR-MOTUS reconstruction was $6.7 \times 6.7 \times 6.7$ mm$^3$, the 2-mm relative shift in the LR direction could be due to a sub-voxel reconstruction offset. For MR-MOTUS, general amplitude variations/discontinuities in the LR direction were also observed at several temporal sections, especially from around 99 s. Such variations are likely due to the batch-based reconstruction of MR-MOTUS to address the memory limits (each batch has around 33 s of data), which yielded slightly different low-rank DVF bases across batches that might affect the cross-batch motion amplitude consistency.

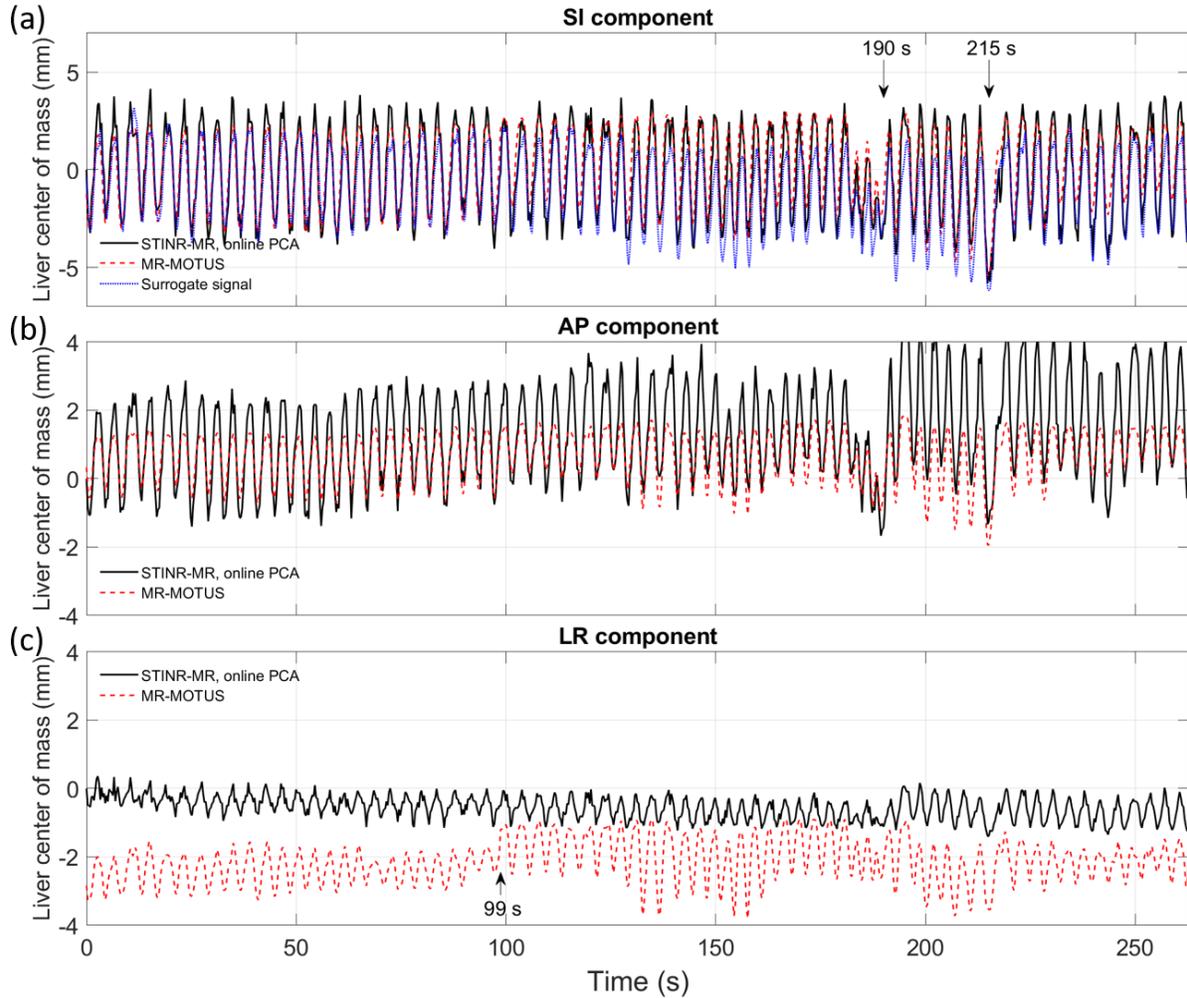

**Figure 10.** Comparison of the liver center-of-mass motion estimated by STINR-MR and MR-MOTUS. Subfigures 10(a-c) show the SI, AP, and LR components of the liver center-of-mass motion, respectively. For comparison, the surrogate signal extracted from the k-space origin was plotted in subfig. 10(a).

## 4. Discussion

In this study, we proposed a joint reconstruction and deformable registration-based framework, STINR-MR, for 3D cine-MRI reconstruction. STINR-MR used a spatial INR and a temporal INR, together with a PCA-based motion model, to reconstruct 3D cine MR images with superior spatial and temporal resolutions. STINR-MR decoupled the challenging spatiotemporal inverse problem into the joint training of two INR networks to separately capture the spatial information and the temporal motion, which allows high-quality dynamic image reconstruction from significantly undersampled data via a 'one-shot' learning scheme. The introduction of the PCA-based motion model helps to regularize the motion fields and reduce the corresponding solution space, and allows the STINR-MR framework to capture highly irregular motion patterns (Fig. 6, Table 3, Table 4, and Supplementary file: Fig. S-2).

4.1 Performance of STINR-MR and online vs. offline PCA models

STINR-MR is generally robust to the simulated motion and anatomical variations (Figs. 6-8, Tables 3-8), for both the online and the offline PCA-based variants. Comparing online and offline PCA, the latter generally provides better results, which is expected as the offline PCA is close to the 'perfect' motion model as long as the underlying anatomy remains similar between the prior 4D-MRI and the new cine MR acquisitions. In contrast, the online PCA has to be built on-the-fly from the cine MR acquisitions and is susceptible to the irregularity of the online motion. Of the six motion scenarios (Fig. 5, Table 2), the two scenarios with substantial baseline shifts, S2 and S3, are more challenging for STINR-MR using an online PCA motion model. The relatively worse performance is expected as the baseline shifts will lead to prominent intra-phase motion artifacts in the reconstructed online 4D-MRI (supplementary materials: Fig. S-1), and these artifacts will propagate into the corresponding online PCA motion model to reduce the accuracy of STINR-MR.

Although the offline PCA shows generally more favorable results compared to online PCA, the differences are not substantial, especially in terms of the tumor localization accuracy (Table 4, Table 6, Table 8, supplementary materials: Fig. S-2). For the inter-scan tumor shrinkage scenario (Table 6), the offline PCA performs slightly worse than the online PCA, as the underlying anatomical change impacts the validity of the offline PCA motion model that is built from prior offline 4D-MR images. In addition, in this study, we did not consider and simulate artifacts in the offline 4D-MRI, which can be caused by undersampling and motion irregularities in real clinical practices. The artifacts in the offline 4D-MRI will similarly propagate into the offline PCA motion model, as for the case of online 4D-MRI, and impact the results of the STINR-MR based on the offline PCA model. We expect the performance gap to be even smaller between online and offline PCA-based models when such scenarios are considered, and future comprehensive studies using more real patient data are warranted. In addition, instead of using PCA to explicitly extracting motion models from online or offline 4D-MRI, we may incorporate another INR to directly learn the motion model from the available data to mitigate the impacts of 4D-MRI artifacts, which remains to be further investigated.

4.2 Simulation and human studies

In this study, we used XCAT simulations to evaluate STINR-MR. One challenge of the XCAT study is to simulate complex-valued MR signals with spatial phase modulation. We adopted a phase modulation simulation strategy used in previous deep learning-based works (Zhu et al. 2018, Terpstra et al. 2020), which shows good generalizability towards real clinical data. Various phase maps were simulated based on the superposition of four sinusoidal oscillations to generate complex-valued MR images with meaningful real and imaginary parts. However, such simulations may not fully represent the phase maps in clinical data. The reference-frame MR images reconstructed in the XCAT study (Fig. 6) appear to have more strip artifacts than those reconstructed in the human subject study (Fig. 9), which could be partially due to the over modulation simulated in phase maps for the XCAT data.

In addition to the XCAT study, we used a human subject dataset to further evaluate STINR-MR. Results show that the motion solved by STINR-MR is similar to that of MR-MOTUS, and to motion surrogate signals directly tracked from the k-space. The human subject study further validates the clinical applicability of STINR-MR, while future investigations are warranted to quantitatively compare STINR-MR to other methods using real patient data to evaluate their accuracy and efficiency. To provide a 'gold-standard' reference to evaluate STINR-MR, we can acquire self-navigation motion surrogate signals interlaced into the pulse sequence, while such evaluation is currently limited to 1D (Huttinga et al. 2022).

For 3D evaluation, anthropomorphic and motion-enabled MR phantoms can be used with well-controlled, 'known' motion. Such a phantom is currently under development in our group for future studies (Chiu et al. 2022).

4.3 Comparison between MR-MOTUS and STINR-MR

Compared with MR-MOTUS, a major advantage of STINR-MR is its ability to fine-tune the reference-frame MR image during the joint image reconstruction and motion solution stage, which helps to remove aliasing and motion artifacts using all k-space data and a simultaneously-optimized motion model. In contrast, MR-MOTUS reconstructed the reference frame prior to the motion solution, and the motion/undersampling artifacts of the reference-frame MR image were propagated and subsequently affected the accuracy of solved intra-scan DVFs. MR-MOTUS was computationally demanding and required substantial memory footprints and computing time. As a result, in our study the whole sequence of MR acquisitions was partitioned into batches for motion estimation to meet the memory constraint. While partitioning can accelerate the reconstruction process and reduce the memory requirements, some discontinuities were observed in the solved motion across the batches (Fig.10 (c)). There was currently no mechanism to enforce consistency and coherence throughout the whole sequence, as the motion estimation between different batches was independent. Such a discontinuity can be mitigated by initializing the basis DVFs of one batch with the ones solved from the previous batch, while such a strategy will cause prolonged reconstruction time (the batches cannot be parallelized) and the errors may propagate from batch to batch. In contrast, STINR-MR is a light-weight and compact framework that can reconstruct the whole sequence of 3D cine-MRI with >1,000 frames in a single shot.

4.4 Limitations

Compared with the XCAT study, we observed an increase of the training time for the human subject study. It was due to the combined effects of increased imaging size, coil number, and the complexity of the underlying anatomy. As the sensitivity profiles of large coil arrays contain redundant information, a coil compression scheme to reduce the number of coils can be used to further accelerate the reconstruction (Buehrer et al. 2007, Huang et al. 2008). In addition, the intra-scan deformation-only assumption of STINR-MR may limit its applicability to dynamic processes involving rapid variations of material constitutions and MR signals, for instance the scenario of perfusion MR imaging. Future update of the STINR-MR framework that allows continuous, non-deformation intensity variations can potentially solve the problem while remains to be investigated.

# 5. Conclusion

STINR-MR presents a joint image reconstruction and deformable registration framework to reconstruct 3D cine-MRI, by using powerful spatial and temporal implicit neural representations with learning-based hash encoding. The results demonstrated that STINR-MR can reconstruct dynamic volumetric MR images of >1,000 frames and <100-ms temporal resolutions per frame, with superior accuracy and efficiency. With its cine-imaging capability, STINR-MR can capture irregular and aperiodic motion patterns and the underlying 3D anatomy to improve MR-guided interventions, such as MR-guided radiotherapy.

**Acknowledgments**

The study was supported by funding from the National Institutes of Health (R01 CA240808, R01 CA258987). We would like to thank Dr. Paul Segars at Duke University for providing the XCAT phantom for the study.

**References**

Asif, M. S., L. Hamilton, M. Brummer and J. Romberg (2013). "Motion-adaptive spatio-temporal regularization for accelerated dynamic MRI." Magn Reson Med **70**(3): 800-812.

Bartsch, A. J., G. Homola, A. Biller, L. Solymosi and M. Bendszus (2006). "Diagnostic functional MRI: illustrated clinical applications and decision-making." J Magn Reson Imaging **23**(6): 921-932.

Bertholet, J., A. Knopf, B. Eiben, J. McClelland, A. Grimwood, E. Harris, M. Menten, P. Poulsen, D. T. Nguyen, P. Keall and U. Oelfke (2019). "Real-time intrafraction motion monitoring in external beam radiotherapy." Physics in Medicine and Biology **64**(15): 15TR01.

Biswas, S., H. K. Aggarwal and M. Jacob (2019). "Dynamic MRI using model-based deep learning and SToRM priors: MoDL-SToRM." Magn Reson Med **82**(1): 485-494.

Brau, A. C. and J. H. Brittain (2006). "Generalized self-navigated motion detection technique: Preliminary investigation in abdominal imaging." Magn Reson Med **55**(2): 263-270.

Buehrer, M., K. P. Pruessmann, P. Boesiger and S. Kozerke (2007). "Array compression for MRI with large coil arrays." Magn Reson Med **57**(6): 1131-1139.

Burdumy, M., L. Traser, F. Burk, B. Richter, M. Echternach, J. G. Korvink, J. Hennig and M. Zaitsev (2017). "One-second MRI of a three-dimensional vocal tract to measure dynamic articulator modifications." J Magn Reson Imaging **46**(1): 94-101.

Chan, R. W., E. A. Ramsay, C. H. Cunningham and D. B. Plewes (2009). "Temporal stability of adaptive 3D radial MRI using multidimensional golden means." Magn Reson Med **61**(2): 354-363.

Chen, Z. Y., Y. B. Chen, J. W. Liu, X. Q. Xu, V. Goel, Z. Y. Wang, H. Shi and X. L. Wang (2022). "VideoINR: Learning Video Implicit Neural Representation for Continuous Space-Time Super-Resolution." 2022 Ieee/Cvf Conference on Computer Vision and Pattern Recognition (Cvpr 2022): 2037-2047.

Chiu, T., S. Ho, J. Visak, M. Willis and Y. Zhang (2022). "Developing a Pneumatic-Driven, Dual-Modal (MR/CT) and Anthropomorphic Breathing Phantom for Image-Guided Radiotherapy." Medical Physics **49**(6): E513-E513.

Cleary, K. and T. M. Peters (2010). "Image-guided interventions: technology review and clinical applications." Annu Rev Biomed Eng **12**: 119-142.

Constantine, G., K. Shan, S. D. Flamm and M. U. Sivananthan (2004). "Role of MRI in clinical cardiology." Lancet **363**(9427): 2162-2171.

Corradini, S., F. Alongi, N. Andratschke, C. Belka, L. Boldrini, F. Cellini, J. Debus, M. Guckenberger, J. Horner-Rieber, F. J. Lagerwaard, R. Mazzola, M. A. Palacios, M. E. P. Philippens, C. P. J. Raaijmakers, C. H. J. Terhaard, V. Valentini and M. Niyazi (2019). "MR-guidance in clinical reality: current treatment challenges and future perspectives." Radiat Oncol **14**(1): 92.

Curtis, A. D. and H. M. Cheng (2022). "Primer and Historical Review on Rapid Cardiac CINE MRI." J Magn Reson Imaging **55**(2): 373-388.

Deng, Z. X., J. N. Pang, W. S. Yang, Y. Yue, B. Sharif, R. Tuli, D. B. Li, B. Fraass and Z. Y. Fan (2016). "Four-Dimensional MRI Using Three-Dimensional Radial Sampling with Respiratory Self-Gating to Characterize

Temporal Phase-Resolved Respiratory Motion in the Abdomen." <u>Magnetic Resonance in Medicine</u> **75**(4): 1574-1585.

Dregely, I., D. Prezzi, C. Kelly-Morland, E. Roccia, R. Neji and V. Goh (2018). "Imaging biomarkers in oncology: Basics and application to MRI." <u>Journal of Magnetic Resonance Imaging</u> **48**(1): 13-26.

Feng, L. (2022). "Golden-Angle Radial MRI: Basics, Advances, and Applications." <u>J Magn Reson Imaging</u> **56**(1): 45-62.

Feng, L., L. Axel, H. Chandarana, K. T. Block, D. K. Sodickson and R. Otazo (2016). "XD-GRASP: Golden-angle radial MRI with reconstruction of extra motion-state dimensions using compressed sensing." <u>Magn Reson Med</u> **75**(2): 775-788.

Feng, L., T. Benkert, K. T. Block, D. K. Sodickson, R. Otazo and H. Chandarana (2017). "Compressed sensing for body MRI." <u>J Magn Reson Imaging</u> **45**(4): 966-987.

Ferzli, R. and L. J. Karam (2005). "No-reference objective wavelet based noise immune image sharpness metric." <u>2005 International Conference on Image Processing (ICIP), Vols 1-5</u>: 1157-1160.

Fessler, J. A. (2010). "Model-Based Image Reconstruction for Mri." <u>IEEE Signal Process Mag</u> **27**(4): 81-89.

Frisoni, G. B., N. C. Fox, C. R. Jack, P. Scheltens and P. M. Thompson (2010). "The clinical use of structural MRI in Alzheimer disease." <u>Nature Reviews Neurology</u> **6**(2): 67-77.

Full, P. M., F. Isensee, P. F. Jäger and K. Maier-Hein (2021). "Studying Robustness of Semantic Segmentation Under Domain Shift in Cardiac MRI." <u>STACOM 2020: Statistical Atlases and Computational Models of the Heart. M&Ms and EMIDEC Challenges</u> **12592**: 238–249.

Grimm, R., S. Furst, M. Souvatzoglou, C. Forman, J. Hutter, I. Dregely, S. I. Ziegler, B. Kiefer, J. Hornegger, K. T. Block and S. G. Nekolla (2015). "Self-gated MRI motion modeling for respiratory motion compensation in integrated PET/MRI." <u>Med Image Anal</u> **19**(1): 110-120.

Hall, W. A., E. S. Paulson, U. A. van der Heide, C. D. Fuller, B. W. Raaymakers, J. J. W. Lagendijk, X. A. Li, D. A. Jaffray, L. A. Dawson, B. Erickson, M. Verheij, K. J. Harrington, A. Sahgal, P. Lee, P. J. Parikh, M. F. Bassetti, C. G. Robinson, B. D. Minsky, A. Choudhury, R. Tersteeg, C. J. Schultz, M. R. L. A. Consortium and C. T. R. C. the ViewRay (2019). "The transformation of radiation oncology using real-time magnetic resonance guidance: A review." <u>Eur J Cancer</u> **122**: 42-52.

Hamilton, J., D. Franson and N. Seiberlich (2017). "Recent advances in parallel imaging for MRI." <u>Prog Nucl Magn Reson Spectrosc</u> **101**: 71-95.

Hansen, M. S. and P. Kellman (2015). "Image reconstruction: an overview for clinicians." <u>J Magn Reson Imaging</u> **41**(3): 573-585.

Heerkens, H. D., M. van Vulpen, C. A. van den Berg, R. H. Tijssen, S. P. Crijns, I. Q. Molenaar, H. C. van Santvoort, O. Reerink and G. J. Meijer (2014). "MRI-based tumor motion characterization and gating schemes for radiation therapy of pancreatic cancer." <u>Radiother Oncol</u> **111**(2): 252-257.

Hornik, K., M. Stinchcombe and H. White (1989). "Multilayer Feedforward Networks Are Universal Approximators." <u>Neural Networks</u> **2**(5): 359-366.

Huang, F., S. Vijayakumar, Y. Li, S. Hertel and G. R. Duensing (2008). "A software channel compression technique for faster reconstruction with many channels." <u>Magn Reson Imaging</u> **26**(1): 133-141.

Huang, Q., Y. Xian, D. Yang, H. Qu, J. Yi, P. Wu and D. N. Metaxas (2021). "Dynamic MRI reconstruction with end-to-end motion-guided network." <u>Med Image Anal</u> **68**: 101901.

Huttinga, N. R. F., T. Bruijnen, C. A. T. van den Berg and A. Sbrizzi (2021). "Nonrigid 3D motion estimation at high temporal resolution from prospectively undersampled k-space data using low-rank MR-MOTUS." <u>Magnetic Resonance in Medicine</u> **85**(4): 2309-2326.

Huttinga, N. R. F., T. Bruijnen, C. A. T. Van Den Berg and A. Sbrizzi (2022). "Real-Time Non-Rigid 3D Respiratory Motion Estimation for MR-Guided Radiotherapy Using MR-MOTUS." <u>IEEE Trans Med Imaging</u> **41**(2): 332-346.

Huttinga, N. R. F., C. A. T. van den Berg, P. R. Luijten and A. Sbrizzi (2020). "MR-MOTUS: model-based non-rigid motion estimation for MR-guided radiotherapy using a reference image and minimal k-space data." Physics in Medicine and Biology **65**(1).

Jahng, G. H., K. L. Li, L. Ostergaard and F. Calamante (2014). "Perfusion magnetic resonance imaging: a comprehensive update on principles and techniques." Korean J Radiol **15**(5): 554-577.

Jaspan, O. N., R. Fleysher and M. L. Lipton (2015). "Compressed sensing MRI: a review of the clinical literature." Br J Radiol **88**(1056): 20150487.

Jung, H., K. Sung, K. S. Nayak, E. Y. Kim and J. C. Ye (2009). "k-t FOCUSS: a general compressed sensing framework for high resolution dynamic MRI." Magn Reson Med **61**(1): 103-116.

Kaiser, J. F. and R. W. Schafer (1980). "On the Use of the I0-Sinh Window for Spectrum Analysis." IEEE Transactions on Acoustics, Speech, and Signal Processing **ASSP-28**(1): 105.

Kelly, C. J., A. Karthikesalingam, M. Suleyman, G. Corrado and D. King (2019). "Key challenges for delivering clinical impact with artificial intelligence." BMC Med **17**(1): 195.

Khan, M. O. and Y. Fang (2022). "Implicit Neural Representations for Medical Imaging Segmentation." Medical Image Computing and Computer Assisted Intervention, Miccai 2022, Pt V **13435**: 433-443.

Klein, S., M. Staring, K. Murphy, M. A. Viergever and J. P. Pluim (2010). "elastix: a toolbox for intensity-based medical image registration." IEEE Trans Med Imaging **29**(1): 196-205.

Larson, A. C., P. Kellman, A. Arai, G. A. Hirsch, E. McVeigh, D. Li and O. P. Simonetti (2005). "Preliminary investigation of respiratory self-gating for free-breathing segmented cine MRI." Magn Reson Med **53**(1): 159-168.

Lever, F. M., I. M. Lips, S. P. Crijns, O. Reerink, A. L. van Lier, M. A. Moerland, M. van Vulpen and G. J. Meijer (2014). "Quantification of esophageal tumor motion on cine-magnetic resonance imaging." Int J Radiat Oncol Biol Phys **88**(2): 419-424.

Li, R. J., J. H. Lewis, X. Jia, T. Y. Zhao, W. F. Liu, S. Wuenschel, J. Lamb, D. S. Yang, D. A. Low and S. B. Jiang (2011). "On a PCA-based lung motion model." Physics in Medicine and Biology **56**(18): 6009-6030.

Li, X. A., C. Stepaniak and E. Gore (2006). "Technical and dosimetric aspects of respiratory gating using a pressure-sensor motion monitoring system." Med Phys **33**(1): 145-154.

Liang, D., J. Cheng, Z. Ke and L. Ying (2020). "Deep Magnetic Resonance Image Reconstruction: Inverse Problems Meet Neural Networks." IEEE Signal Process Mag **37**(1): 141-151.

Liao, J. R., J. M. Pauly, T. J. Brosnan and N. J. Pelc (1997). "Reduction of motion artifacts in cine MRI using variable-density spiral trajectories." Magn Reson Med **37**(4): 569-575.

Lingala, S. G., B. P. Sutton, M. E. Miquel and K. S. Nayak (2016). "Recommendations for real-time speech MRI." J Magn Reson Imaging **43**(1): 28-44.

Lustig, M., D. Donoho and J. M. Pauly (2007). "Sparse MRI: The application of compressed sensing for rapid MR imaging." Magn Reson Med **58**(6): 1182-1195.

Mildenhall, B., P. P. Srinivasan, M. Tancik, J. T. Barron, R. Ramamoorthi and R. Ng (2022). "NeRF: Representing Scenes as Neural Radiance Fields for View Synthesis." Communications of the Acm **65**(1): 99-106.

Molaei, A., A. Aminimehr, A. Tavakoli, A. Kazerouni, B. Azad, R. Azad and D. Merhof (2023). "Implicit Neural Representation in Medical Imaging: A Comparative Survey." arXiv preprint arXiv:2307.16142.

Muckley, M. J., R. Stern, T. Murrell and F. Knoll (2020). TorchKbNufft: A High-Level, Hardware-Agnostic Non-Uniform Fast Fourier Transform. ISMRM Workshop on Data Sampling & Image Reconstruction.

Muller, T., A. Evans, C. Schied and A. Keller (2022). "Instant Neural Graphics Primitives with a Multiresolution Hash Encoding." Acm Transactions on Graphics **41**(4).

Nayak, K. S., Y. Lim, A. E. Campbell-Washburn and J. Steeden (2022). "Real-Time Magnetic Resonance Imaging." J Magn Reson Imaging **55**(1): 81-99.

Padilla, L., A. Havnen-Smith, L. Cervino and H. A. Al-Hallaq (2019). "A survey of surface imaging use in radiation oncology in the United States." Journal of Applied Clinical Medical Physics **20**(12): 70-77.

Petralia, G., P. E. Summers, A. Agostini, R. Ambrosini, R. Cianci, G. Cristel, L. Calistri and S. Colagrande (2020). "Dynamic contrast-enhanced MRI in oncology: how we do it." Radiologia Medica **125**(12): 1288-1300.

Pollard, J. M., Z. F. Wen, R. Sadagopan, J. H. Wang and G. S. Ibbott (2017). "The future of image-guided radiotherapy will be MR guided." British Journal of Radiology **90**(1073).

Rajiah, P. S., C. J. Francois and T. Leiner (2023). "Cardiac MRI: State of the Art." Radiology **307**(3): e223008.

Rao, C. L., Q. Wu, P. Q. Zhou, J. Y. Yu, Y. Y. Zhang and X. Lou (2023). "An Energy-Efficient Accelerator for Medical Image Reconstruction From Implicit Neural Representation." Ieee Transactions on Circuits and Systems I-Regular Papers **70**(4): 1625-1638.

Ravishankar, S., J. C. Ye and J. A. Fessler (2020). "Image Reconstruction: From Sparsity to Data-adaptive Methods and Machine Learning." Proc IEEE Inst Electr Electron Eng **108**(1): 86-109.

Reed, A. W., H. Kim, R. Anirudh, K. A. Mohan, K. Champley, J. G. Kang and S. Jayasuriya (2021). "Dynamic CT Reconstruction from Limited Views with Implicit Neural Representations and Parametric Motion Fields." 2021 Ieee/Cvf International Conference on Computer Vision (Iccv 2021): 2238-2248.

Rudin, L. I., S. Osher and E. Fatemi (1992). "Nonlinear Total Variation Based Noise Removal Algorithms." Physica D **60**(1-4): 259-268.

Schlemper, J., J. Caballero, J. V. Hajnal, A. N. Price and D. Rueckert (2018). "A Deep Cascade of Convolutional Neural Networks for Dynamic MR Image Reconstruction." Ieee Transactions on Medical Imaging **37**(2): 491-503.

Segars, W. P., G. Sturgeon, S. Mendonca, J. Grimes and B. M. Tsui (2010). "4D XCAT phantom for multimodality imaging research." Med Phys **37**(9): 4902-4915.

Shao, H. C., T. Li, M. J. Dohopolski, J. Wang, J. Cai, J. Tan, K. Wang and Y. Zhang (2022). "Real-time MRI motion estimation through an unsupervised k-space-driven deformable registration network (KS-RegNet)." Physics in Medicine and Biology **67**(13).

Shen, L. Y., J. Pauly and L. Xing (2022). "NeRP: Implicit Neural Representation Learning With Prior Embedding for Sparsely Sampled Image Reconstruction." Ieee Transactions on Neural Networks and Learning Systems.

Sitzmann, V., J. Martel, A. Bergman, D. Lindell and G. Wetzstein (2020). "Implicit neural representations with periodic activation functions." Advances in Neural Information Processing Systems **33**: 7462-7473.

Sourbron, S. P. and D. L. Buckley (2013). "Classic models for dynamic contrast-enhanced MRI." NMR Biomed **26**(8): 1004-1027.

Stemkens, B., E. S. Paulson and R. H. N. Tijssen (2018). "Nuts and bolts of 4D-MRI for radiotherapy." Phys Med Biol **63**(21): 21TR01.

Tancik, M., P. P. Srinivasan, B. Mildenhall, S. Fridovich-Keil, N. Raghavan, U. Singhal, R. Ramamoorthi, J. T. Barron and R. Ng (2020). "Fourier Features Let Networks Learn High Frequency Functions in Low Dimensional Domains." NeurIPS.

Terpstra, M. L., M. Maspero, T. Bruijnen, J. J. C. Verhoeff, J. J. W. Lagendijk and C. A. T. van den Berg (2021). "Real-time 3D motion estimation from undersampled MRI using multi-resolution neural networks." Med Phys **48**(11): 6597-6613.

Terpstra, M. L., M. Maspero, F. d'Agata, B. Stemkens, M. P. W. Intven, J. J. W. Lagendijk, C. A. T. van den Berg and R. H. N. Tijssen (2020). "Deep learning-based image reconstruction and motion estimation from undersampled radial k-space for real-time MRI-guided radiotherapy." Physics in Medicine and Biology **65**(15).

Tewari, A., J. Thies, B. Mildenhall, P. Srinivasan, E. Tretschk, W. Yifan, C. Lassner, V. Sitzmann, R. Martin-Brualla, S. Lombardi, T. Simon, C. Theobalt, M. Niessner, J. T. Barron, G. Wetzstein, M. Zollhofer and V. Golyanik (2022). "Advances in Neural Rendering." Computer Graphics Forum **41**(2): 703-735.

Tsao, J., P. Boesiger and K. P. Pruessmann (2003). "k-t BLAST and k-t SENSE: dynamic MRI with high frame rate exploiting spatiotemporal correlations." Magn Reson Med **50**(5): 1031-1042.

Uecker, M., S. Zhang and J. Frahm (2010). "Nonlinear inverse reconstruction for real-time MRI of the human heart using undersampled radial FLASH." Magn Reson Med **63**(6): 1456-1462.

Vedam, S. S., P. J. Keall, V. R. Kini and R. Mohan (2001). "Determining parameters for respiration-gated radiotherapy." Med Phys **28**(10): 2139-2146.

Wild, J. M., M. N. Paley, L. Kasuboski, A. Swift, S. Fichele, N. Woodhouse, P. D. Griffiths and E. J. van Beek (2003). "Dynamic radial projection MRI of inhaled hyperpolarized 3He gas." Magn Reson Med **49**(6): 991-997.

Winkelmann, S., T. Schaeffter, T. Koehler, H. Eggers and O. Doessel (2007). "An optimal radial profile order based on the Golden Ratio for time-resolved MRI." IEEE Trans Med Imaging **26**(1): 68-76.

Witt, J. S., S. A. Rosenberg and M. F. Bassetti (2020). "MRI-guided adaptive radiotherapy for liver tumours: visualising the future." Lancet Oncol **21**(2): e74-e82.

Yasue, K., H. Fuse, S. Oyama, K. Hanada, K. Shinoda, H. Ikoma, T. Fujisaki and Y. Tamaki (2022). "Quantitative analysis of the intra-beam respiratory motion with baseline drift for respiratory-gating lung stereotactic body radiation therapy." J Radiat Res **63**(1): 137-147.

Zech, J. R., M. A. Badgeley, M. Liu, A. B. Costa, J. J. Titano and E. K. Oermann (2018). "Variable generalization performance of a deep learning model to detect pneumonia in chest radiographs: A cross-sectional study." PLoS Med **15**(11): e1002683.

Zha, R. Y., Y. H. Zhang and H. D. Li (2022). "NAF: Neural Attenuation Fields for Sparse-View CBCT Reconstruction." Medical Image Computing and Computer Assisted Intervention, Miccai 2022, Pt Vi **13436**: 442-452.

Zhang, Y., M. R. Folkert, X. Huang, L. Ren, J. Meyer, J. N. Tehrani, R. Reynolds and J. Wang (2019). "Enhancing liver tumor localization accuracy by prior-knowledge-guided motion modeling and a biomechanical model." Quant Imaging Med Surg **9**(7): 1337-1349.

Zhang, Y., J. Ma, P. Iyengar, Y. Zhong and J. Wang (2017). "A new CT reconstruction technique using adaptive deformation recovery and intensity correction (ADRIC)." Med Phys **44**(6): 2223-2241.

Zhang, Y., H. C. Shao, T. Pan and T. Mengke (2023). "Dynamic cone-beam CT reconstruction using spatial and temporal implicit neural representation learning (STINR)." Phys Med Biol **68**(4).

Zhang, Y., F. F. Yin, W. P. Segars and L. Ren (2013). "A technique for estimating 4D-CBCT using prior knowledge and limited-angle projections." Medical Physics **40**(12): 121701.

Zhao, B., J. P. Haldar, A. G. Christodoulou and Z. P. Liang (2012). "Image reconstruction from highly undersampled (k, t)-space data with joint partial separability and sparsity constraints." IEEE Trans Med Imaging **31**(9): 1809-1820.

Zhu, B., J. Z. Liu, S. F. Cauley, B. R. Rosen and M. S. Rosen (2018). "Image reconstruction by domain-transform manifold learning." Nature **555**(7697): 487-492.

# Supplementary materials

## 1. Initial reference-frame images via NUFFT reconstruction for the progressive training scheme

Figure S-1 presents the initial reference-frame MR images reconstructed by NUFFT, which served as the training target in the first stage of the progressive training scheme (Sec. 2.3.3). The NUFFT-reconstructed reference-frame images contained undersampling and motion artifacts, which were gradually corrected via further trainings in Stage 2 and Stage 3.

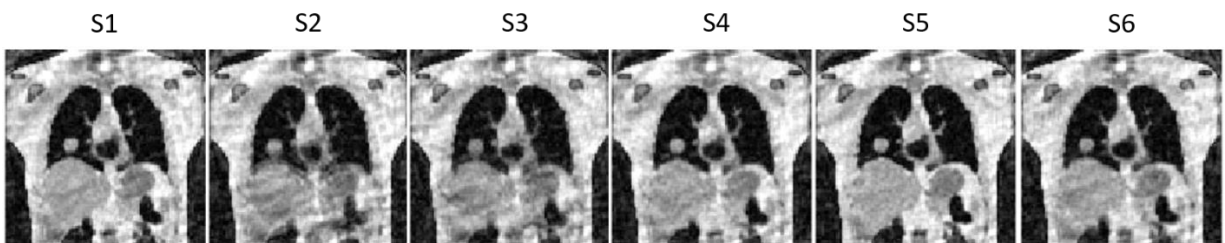

**Figure S-1.** Initial reference-frame MR images reconstructed by NUFFT for motion scenarios S1-S6 under the progressive training scheme.

## 2. Additional results of the XCAT phantom study

Figure S-2 compares the lung tumor center-of-mass motion solved by different methods in the superior-inferior (SI) direction for the six motion scenarios. Overall, all methods captured the motion variations, but MR-MOTUS presented the worst localization accuracy. Overshoots and undershoots can be observed at the peaks and troughs of the motion curves solved by MR-MOTUS.

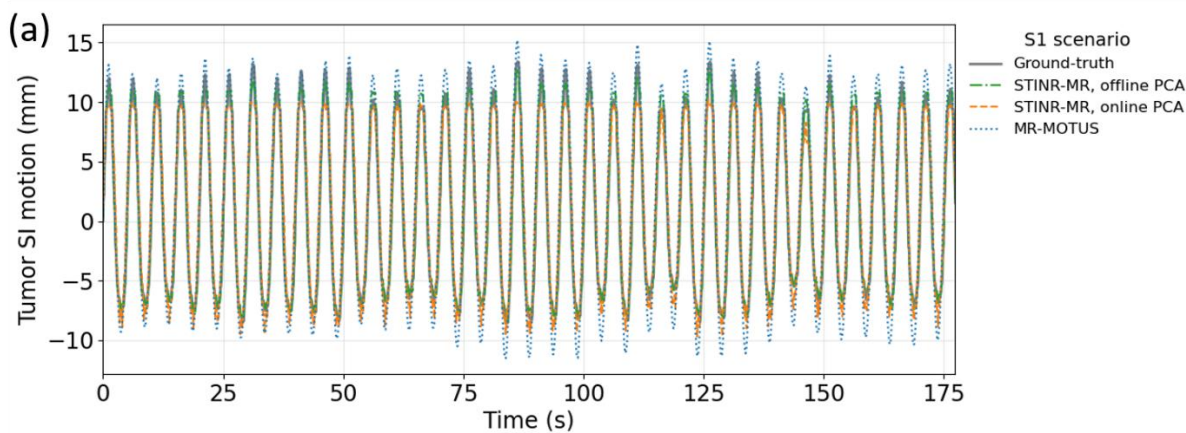

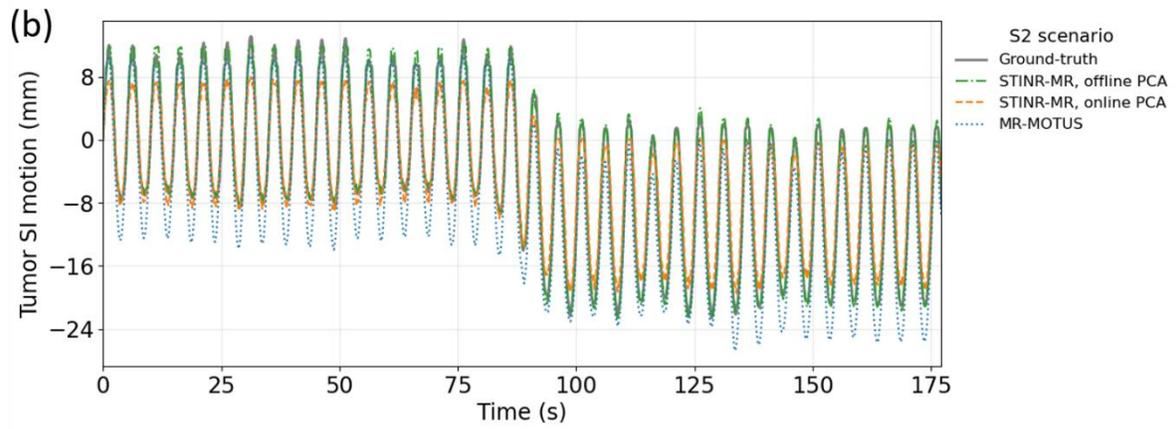

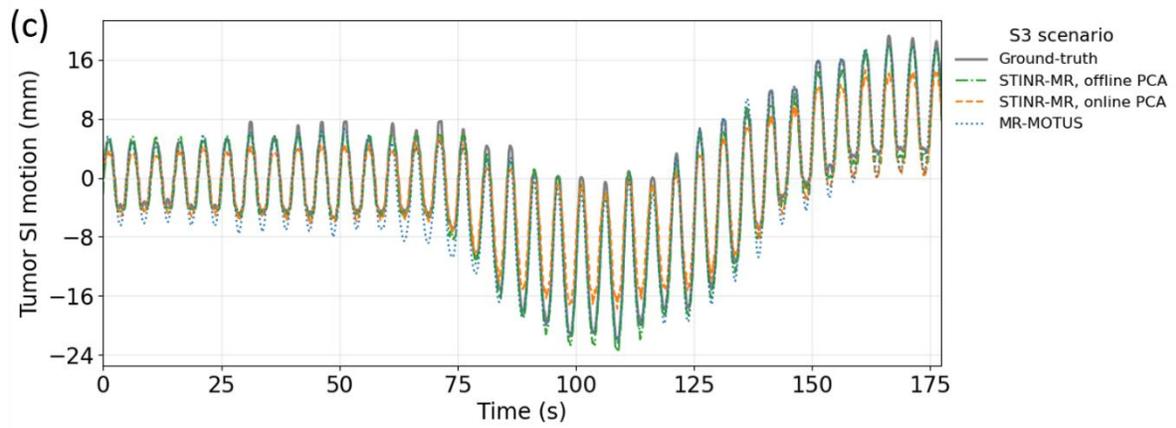

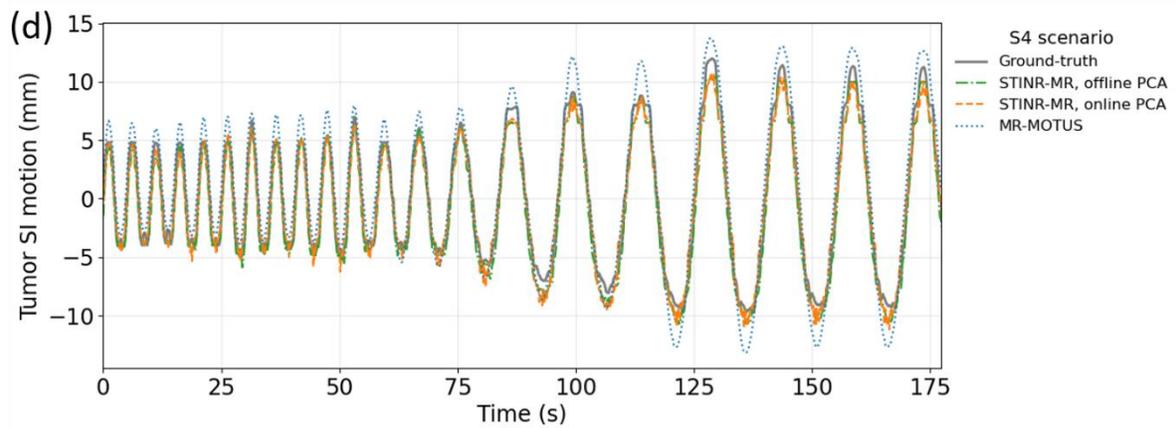

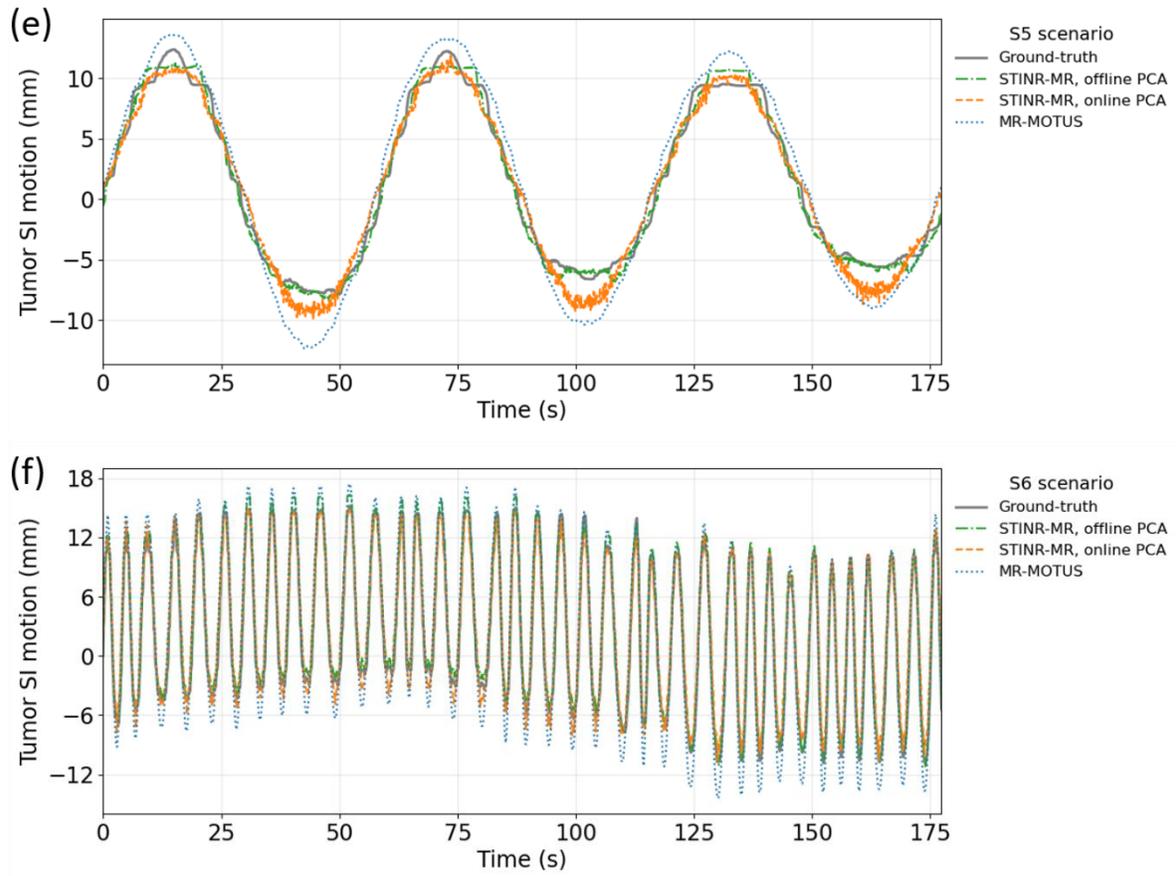

**Figure S-2.** Comparison of the lung tumor center-of-mass motion solved by different methods in the superior-inferior (SI) direction for the six motion scenarios (S1-S6).